\documentclass[aps,superscriptaddress,amsmath,amssymb,floatfix,twocolumn,showpacs,amsfonts,longbibliography,pra]{revtex4-1}
\usepackage{times}
\usepackage[varg]{txfonts}
\usepackage{textcomp}
\usepackage{graphicx}
\usepackage{subfigure}
\usepackage{tabu}
\usepackage{color}
\usepackage[colorlinks=true,citecolor=blue,urlcolor=blue,linkcolor=blue,hyperindex]{hyperref}
\usepackage{braket}
\usepackage{float}
\usepackage{overpic}
\usepackage{gensymb}
\usepackage{mathrsfs}
\usepackage{tikz}
\usepackage{bm}
\usepackage{multirow}
\usepackage{amsfonts}
\usepackage{amsmath}
\usepackage{mathrsfs}
\usepackage{amssymb}
\usepackage{paralist}
\usepackage{indentfirst}
\usepackage{ulem}
\usepackage{soul}
\usepackage{cancel}
\usepackage{CJK}

\newcommand{\dima}{{\tikz[scale=0.8]{\draw[line width=1.5pt] (0,-0.2) -- (0.1732,-0.1) (0.1732, 0.1)  -- (0, 0.2)   (-0.1732,0.1)  -- (-0.1732,-0.1) ;\draw[line width=0.5pt,densely dotted]   (0.1732,-0.1) -- (0.1732, 0.1)  (0, 0.2) --  (-0.1732,0.1) (-0.1732,-0.1) --  (0,-0.2);}}}
\newcommand{\dimb}{{\tikz[scale=0.8]{\draw[line width=0.5pt,densely dotted] (0,-0.2) -- (0.1732,-0.1) (0.1732, 0.1)  -- (0, 0.2)   (-0.1732,0.1)  -- (-0.1732,-0.1) ;\draw[line width=1.5pt]   (0.1732,-0.1) -- (0.1732, 0.1)  (0, 0.2) --  (-0.1732,0.1) (-0.1732,-0.1) --  (0,-0.2);}}}
\newcommand{\hexa}{{\tikz[scale=1.0]{\fill (0,-0.1) circle (0.5mm);\fill (0,0.2464) circle (0.5mm);\fill (0.3,0.0732) circle (0.5mm);\fill (0.1,0.0732) circle (0.5mm);\draw (0,-0.1) -- (-0.1,0.0732) -- (0,0.2464) -- (0.2,0.2464) -- (0.3,0.0732) -- (0.2,-0.1) -- (0,-0.1);\draw (0,-0.1) -- (0.2,0.2464) (0.2,-0.1) -- (0, 0.2464) (-0.1,0.0732) -- (0.3,0.0732);}}}
\newcommand{\hexb}{{\tikz[scale=1.0]{\fill (0,-0.1) circle (0.5mm);\fill (0,0.2464) circle (0.5mm);\fill (0.3,0.0732) circle (0.5mm);\draw (0,-0.1) -- (-0.1,0.0732) -- (0,0.2464) -- (0.2,0.2464) -- (0.3,0.0732) -- (0.2,-0.1) -- (0,-0.1);\draw (0,-0.1) -- (0.2,0.2464) (0.2,-0.1) -- (0, 0.2464) (-0.1,0.0732) -- (0.3,0.0732);}}}
\allowdisplaybreaks

\begin{document}
\title{Frustrated Rydberg Atom Arrays Meet Cavity-QED: Emergence of the Superradiant Clock Phase}
\author{Ying Liang}
\affiliation{Department of Physics, and Chongqing Key Laboratory for Strongly Coupled Physics, Chongqing University, Chongqing, 401331, China}

\author{Bao-Yun Dong}
\affiliation{Department of Physics, and Chongqing Key Laboratory for Strongly Coupled Physics, Chongqing University, Chongqing, 401331, China}

\author{Zi-Jian Xiong}
\affiliation{College of Physics and Electronic Engineering, Chongqing Normal University, Chongqing 401331, China}

\author{Xue-Feng Zhang}
\thanks{corresponding author:  zhangxf@cqu.edu.cn}
\affiliation{Department of Physics, and Chongqing Key Laboratory for Strongly Coupled Physics, Chongqing University, Chongqing, 401331, China}
\affiliation{Center of Quantum Materials and Devices, Chongqing University, Chongqing 401331, China}

\begin{abstract}
Rydberg atom triangular arrays in an optical cavity serve as an ideal platform for understanding the interplay between geometric frustration and quantized photons. Using a large-scale quantum Monte Carlo method, we obtain a rich ground state phase diagram. Around half-filling, the infinite long-range light-matter interaction lifts the ground state degeneracy, resulting in a novel order-coexisted superradiant clock phase that completely destroys the fragile order-by-disorder phase observed in classical light fields. According to the Ginzburg-Landau theory, this replacement may result from the competition between threefold and sixfold clock terms. Similar to the spin supersolid, the clear first-order phase transition at the $Z_2$ symmetry line is attributed to the nonzero photon density, which couples to the threefold clock term. Finally, we discuss the low-energy physics in the dimer language and propose that cavity-mediated nonlocal ring exchange interactions may play a critical role in the rich physics induced by the attachment of cavity-QED. Our work opens a new arena of research on the emergent phenomena of quantum phase transitions in many-body quantum optics.
\end{abstract}
\maketitle

\textit{Introduction.--} Matter coupled with light forms a key platform in quantum optics, where infinite long-range interactions give rise to a variety of exotic phenomena \cite{qed-review,long_range}, particularly quantum phase transitions (QPTs) \cite{QPT-rabi, QPT-rabi2, QPT-rabi3}. Recent advancements, driven by the inclusion of additional degrees of freedom, have led to significant progress in understanding QPTs in light-matter quantum few-body systems, including phenomena such as multi-criticality \cite{multi-criticality,multi_criticality} and quantum magnetic clusters \cite{zhangQuantumPhasesQuantum2021, QPT-frust,zyy2}. On the other hand, the integration of quantum many-body systems with cavity-QED \cite{cavity_glassy,cavity_meson,kai,Dalmonte_meson,cavity_pretherm,cavity_spinliquid} has begun to reshape our understanding of emergent phenomena and critical behavior in strongly interacting many-body quantum optics \cite{longrang_shortrange,long_range}.

Quantum frustration has garnered significant attention due to its abundance of exotic emergent phenomena, such as supersolid (SS) \cite{ss_yamamoto,ss_zhang,ss_melko,ss_liwei}, emergent lattice gauge theory \cite{lgt_review,lgt_subir,lgt_zhang1,lgt_zhang2}, quantum spin liquid \cite{sl_zhou,sl_balents,sl_Senthil}, and deconfined criticality \cite{dqcp,dqcp_youjin,dqcp_sandvik,dqcp_yu,dqcp_ZYM,dqcp_zhang,dqcp_xu}. A characteristic example is the antiferromagnetic Ising model on a triangular lattice, where the competition between antiferromagnetic interactions and geometric frustration leads to a disordered ground state with high macroscopic degeneracy \cite{wannierAntiferromagnetismTriangularIsing1950}. Interestingly, even infinitesimal quantum fluctuations can lift this degeneracy and result in an ordered ground state \cite{ss_melko,obd_moessner,obd_moessner2}. This phenomenon, known as the order-by-disorder (OBD) mechanism, is a representative example of the emergent behavior in such systems \cite{fazekasGroundStateProperties1974}.

\begin{figure}[h]
	\centering
	\includegraphics[width=0.9\linewidth]{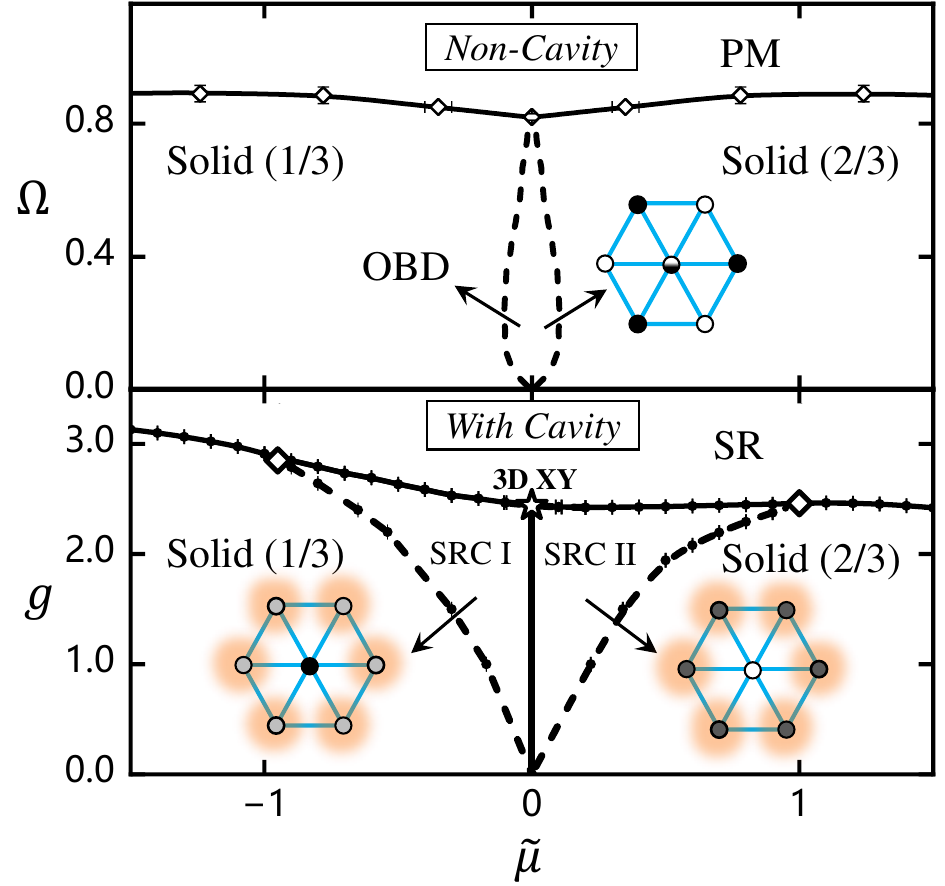}
    \caption{Quantum phase diagrams of Rydberg atom arrays with or without a cavity. \textit{Top}: The non-cavity phase diagram is obtained from Ref.~\cite{liuIntrinsicQuantumIsing2020} by transferring the parameters of the TFIM. \textit{Bottom}: The phase diagram in a cavity is calculated by QMC simulation at $\Delta=9$, where $\tilde{\mu}=\mu_b-3V$. The dashed (solid) lines mark the second (first)-order phase transition. The diamond points represent the two triple points, and the star point labels the 3D XY critical point. The schematic pictures of the OBD phase and SRC phases are pointed with the black arrows.}
    \label{fig1}
\end{figure}

Recent advancements in Rydberg atom arrays have provided an ideal platform for simulating strongly correlated two-level atoms with exceptional tunability \cite{browaeys2020many,Dong2025EngineeringAA}. The strongly interactive Rydberg atoms can be effectively mapped onto the extended transverse field Ising model (TFIM), in which the Rabi frequency $\Omega$ plays as a transverse field \cite{Rydberg_chain,Rydberg_nature1}. By arranging tweezer sites with frustrated geometry, the interplay between frustration and the classical light field exhibits various exotic emergent phenomena, including not only the OBD phase \cite{Rydberg_nature2}, but also the spin liquid \cite{Rydberg_longr}, spin glass \cite{Rydberg_glass1,Rydberg_glass2}, and string breaking \cite{Rydberg_kagome,Rydberg_kagome_string_breaking_xu}. Therefore, coupling a frustrated Rydberg atom array with an optical cavity \cite{cavity_tweezer1,cavity_tweezer2,cavity_tweezer3,cavity_tweezer4} is expected to offer novel insights into emergent physics through the interplay among geometric frustration, strong Rydberg interaction, and the quantized photonic field.

In this manuscript, we investigate the frustrated Rydberg atom arrays coupled with quantized photonic fields by using a large-scale quantum Monte Carlo (QMC) simulation. Distinctively from the classical photonic field, as demonstrated in Fig.~\ref{fig1}, the physics of OBD mechanisms is strongly enriched by the attachment of the cavity. Assisted by the infinite long-range light-matter interaction, a novel order coexisted phase named the superradiant clock (SRC) phase emerges and completely ruins the fragile OBD phase without a cavity. Different from the OBD phase with sixfold clock order, the SRC phase is driven by the threefold clock term reminiscent of the spin SS \cite{ss_troyer05,ss_melko,ss_damle05,ss_prokof,XXZ-energy-gap}. Similar to the SS phases, the condensation also happens in the honeycomb backbone and the QPT to the disordered phase exhibits exotic criticality behaviors. However, in contrast to the mist of QPT between SS phases at zero magnetic field, two SRC phases undergo a clear first-order QPT. It hints the photon-induced particle-hole symmetry breaking strongly enhances the threefold clock term to being relevant. At last, according to the analysis of dimer representation, we think the nonlocal ring exchange interaction may play a critical role in the microscopic mechanism of the SRC phase's emergence.

\textit{Model and Methods.--} The Rydberg atom arrays coupled with cavity-QED can be effectively described by the following Hamiltonian \cite{zhangRydbergPolaritonsCavity2013,anQuantumPhaseTransition2022}
\begin{equation}
    H=\frac{g}{\sqrt{N}}\sum^{N}_{i=1}{\left(b^{\dagger}_ia+a^{\dagger}b_i\right)}+\sum^{N}_{<ij>}V_{ij}{n_i^{(b)}n_j^{(b)}}-\mu n^{(a)}-\mu_b\sum^{N}_{i=1} n_i^{(b)},
    \label{eq1}
\end{equation}
where $a$ denotes the bosonic photon field, $b_i$ is the hardcore bosonic operator describing the Rydberg atom with ground state $|g\rangle$ and excited Rydberg state $|r\rangle$ at the $i$-th site, and the corresponding density operators are $n^{(a)}$ and $n_i^{(b)}$. The atom-photon coupling strength $g$ can be tuned by adjusting the two-photon process. The chemical potential $\mu<0$ and gap energy $\Delta=(\mu_b-\mu)>0$ are related to the photon energy and detuning of the Rydberg atomic energy levels, $N$ is the number of tweezer sites, and the Van der Waals (VdW) interactions between Rydberg states are expressed as $V_{ij}=V/R_{ij}^6$. Here, the VdW interaction is truncated to the nearest neighbor, and the influence of longer interactions will be discussed later.

When the atom-photon coupling is switched off, the Rydberg atoms are decoupled from the cavity, so that the ground state is only governed by the interplay between the VdW interaction and the chemical potential $\mu_b$. At the ${Z}_2$ (or spin-up-down) symmetry line $\tilde{\mu}=\mu_b-3V=0$, to minimize the ground state energy, one or two atoms are allowed to stay in the Rydberg state inside a triangle unit. Such local constraints can result in macroscopic degeneracy in the ground state \cite{wannierAntiferromagnetismTriangularIsing1950}. However, a tiny positive (negative) $\tilde{\mu}$ can lift the degeneracy and drive the system away from half-filling so that the translational symmetry is spontaneously broken and the system enters the solid phase with 2/3 (1/3) filling.

In the large $g$ limit, the strong coupling between two-level atoms and the photon can make them form polaritons. Then, their condensation can drive the system into the superradiant (SR) phase, in which the $U(1)$ symmetry is spontaneously broken. According to the Ginzburg-Landau theory, the quantum phase transition between the solid phase and the SR phase should be first order or there exists an intermediate phase between them, such as the superradiant-solid phase which breaks both symmetries just like the SS \cite{zhangRydbergPolaritonsCavity2013,anQuantumPhaseTransition2022}. Then, it would be interesting to think about the interplay between the quantum photonic field and geometric frustration, both of which can bring in the disorder.

The QPTs concerning the OBD mechanism are usually unconventional, such as the multi-criticality of the SS phase \cite{ss_melko,ss_yamamoto,ss_zhang,ss_liwei}, or the deconfined criticality of valence bond crystal \cite{dqcp_zhang,dqcp_xu}. Therefore, it is necessary to borrow high-performance numerical methods. Here, we utilize an efficient large-scale QMC method \cite{zhangRydbergPolaritonsCavity2013,anQuantumPhaseTransition2022}. The observables are calculated by taking an average of 10$^6$ samples after half a million thermalization steps, with the inverse temperature set to $\beta=10L/3$ which is low enough to capture the zero-temperature physics. Without loss of generality, we set $V$=1 as the energy unit and $\Delta=9$ (the influence of $\Delta$ is discussed in the supplementary material (SM) \cite{sm}).

\begin{figure}[t]
	\centering
	\includegraphics[width=1.0\linewidth]{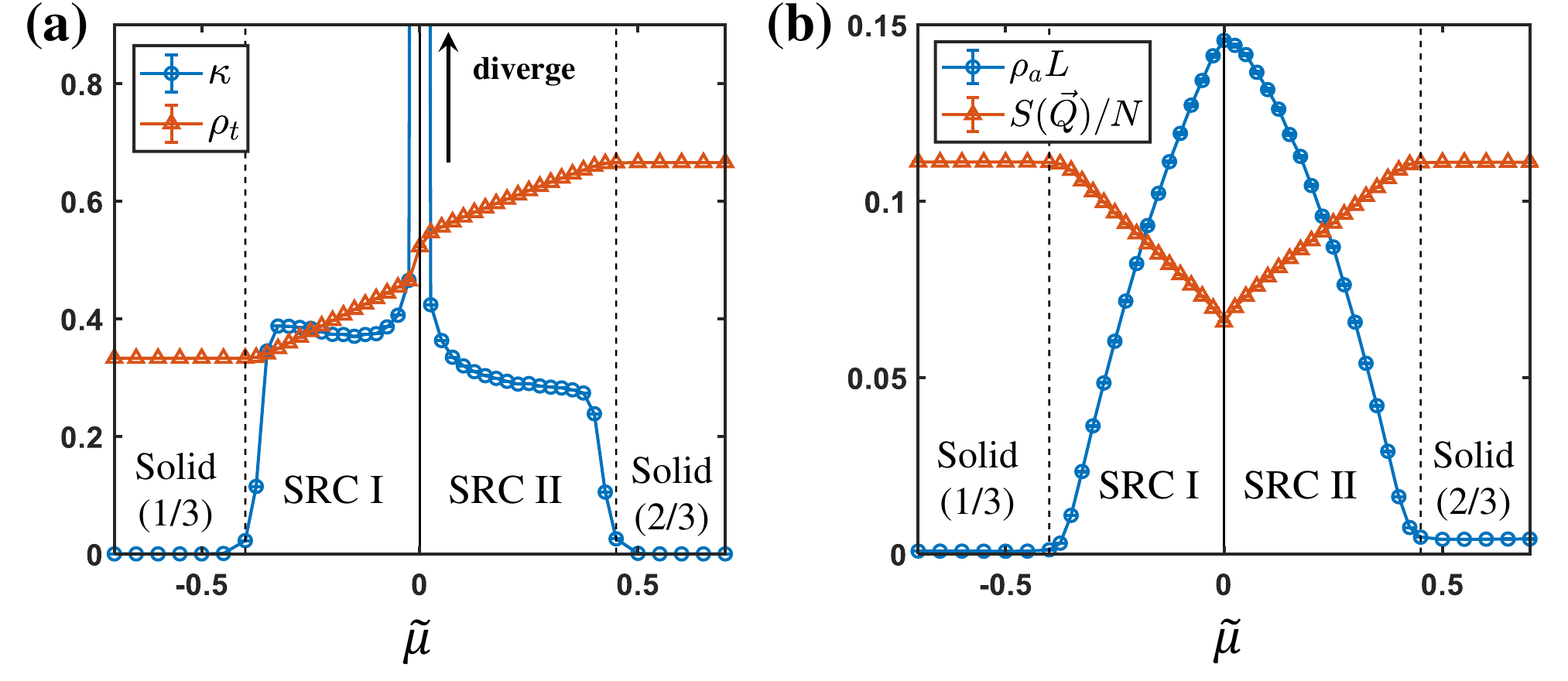}
	\caption{(a) Compressibility $\kappa$ and the average total density $\rho_t$; (b) photon density $\rho_a$  and the structure factor $S(\vec{Q})/N$, calculated via QMC simulation at $g=1.8$ and $L=24$. 
}\label{fig2}
\end{figure}

\textit{Phase Diagram.--} In contrast to the classical light field \cite{Rydberg_nature2}, the coupling between the quantized photonic field and two energy levels imposes an additional U(1) symmetry, causing the conservation of the total density $N_t=\langle n^{(a)}+\sum_i n^{(b)}_i\rangle$. Meanwhile, the large energy gap due to the Rydberg interaction makes the trivial solid phases incompressible. Therefore, as shown in Fig.~\ref{fig2} (a), the solid phases can be identified by the plateaus of the average total density $\rho_t=N_t/N$ and zero compressibility $\kappa=N\beta\left(\langle\rho_t^2\rangle-\langle\rho_t\rangle^2\right)$. Certainly, as translational symmetry is spontaneously broken in both the 1/3 and 2/3 solids, they exhibit long-range diagonal correlations, reflected by the nonzero structure factor $S(\vec{Q})=|s(\vec{Q})|^2$  ($\vec{Q}=(4\pi/3,0)$)  where $s(\vec{Q})=\langle\sum{n_l^{(b)}e^{i\vec{Q}\cdot\vec{R_l}}}/\sqrt{N}\rangle$, as shown in Fig.~\ref{fig2} (b). Meanwhile, the energy gap prevents the formation of polaritons, so the photon density $\rho_a=\langle n^{(a)} \rangle/N$ is nearly zero. The slight deviations of $\rho_a$ and $S(\vec{Q})/N$ from their classical limit values $0$ and $1/9$ can be easily interpreted as second-order perturbations caused by local quantum fluctuations.

Between the two solid phases, the SRC phase can be clearly identified by the nonzero values of both the structure factor and photon density, as shown in Fig.~\ref{fig2}(b). The smooth variation of these two order parameters indicates that the phase transition from the SRC phase to the solid phase is of the second order. Analogous to the SS phases in the triangular lattice \cite{ss_zhang01,ss_zhang02,ss_zhang,ss_melko}, the emergence of the SRC I (II) phase results from the melting of the 1/3 (2/3) solid phase, triggered by inserting particle (hole) excitations on the honeycomb backbone filled with holes (particles). However, unlike the superfluid flowing on the honeycomb backbone in the SS phase, the spontaneous U(1) symmetry breaking in the SRC phase is due to the effective infinite long-range interaction among polaritons teleporting on the honeycomb backbone. We have also calculated the quantum phase boundaries analytically using strong coupling expansion methods \cite{ss_zhang02,zhangRydbergPolaritonsCavity2013}, and they match very well with the numerical results (see SM \cite{sm}). Following this scenario, if we examine the average local magnetization $m=n^{(b)}-1/2$ in the three sublattices, the corresponding signs in the SRC I and II phases would be $(-,-,+)$ and $(-,+,+)$, respectively (see SM \cite{sm}). Given their strong distinction from the $(-,0,+)$ signature of the OBD phase in the classical light field, the existence of the $(-,0,+)$ phase along the $Z_2$ symmetry line becomes highly challenging.

\begin{figure}[t]
	\centering
	\includegraphics[width=0.99\linewidth]{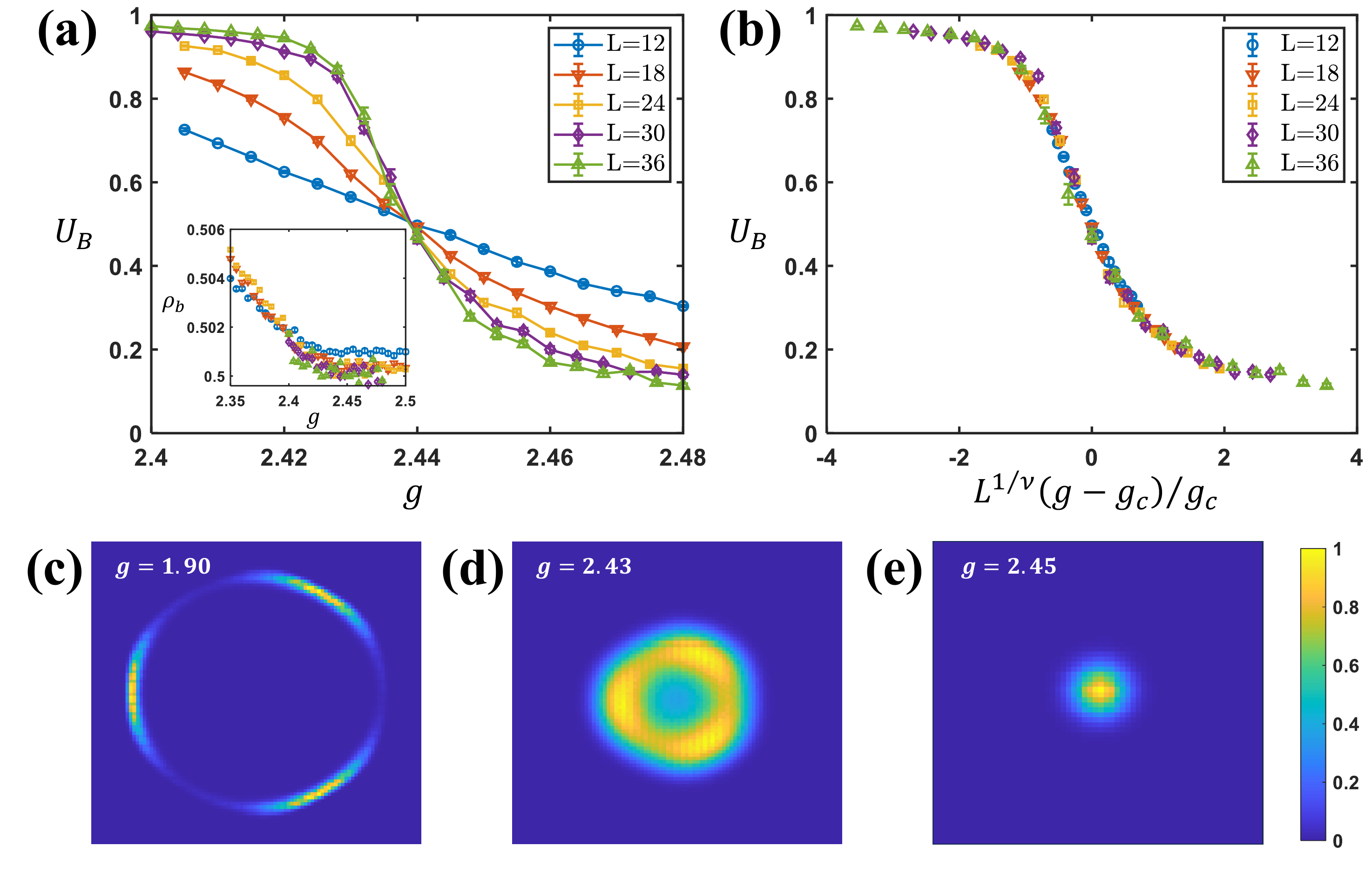}
	\caption{(a) Binder cumulant $U_B$ at $\tilde{\mu}=0$ for different system sizes (inset: Rydberg state occupation density), and corresponding (b) finite-size scaling analysis with data collapse. The extracted critical exponent is $1/\nu=1.50\pm0.09$ with phase transition point $g_c=2.440\pm0.00072$. (c-e) Histogram of $s(\vec{Q})$ in the complex plane for different $g$ at $L=24$.
 }\label{fig4}
\end{figure}

While approaching $\tilde{\mu}=0$, the number of photons continuously increases, indicating the proliferation of polaritons accompanied by a descent in the structure factors. The finite $\kappa$ reflects that the SRC phase is compressible, and the divergence of $\kappa$ with a sudden jump of $\rho_t$ occurring precisely at $\tilde{\mu}=0$ directly points to an obvious first-order QPT, rather than a possible intermediate $(-,0,+)$ phase. However, as shown in Fig.~\ref{fig2}(b), neither the structure factor nor the photon density exhibits any discontinuities. To detect the QPTs around $\tilde{\mu}=0$ in more detail, we further consider the dimensionless quantity: the Binder cumulant of the structure factor, $U_B=2-\langle S(\vec{Q})^2\rangle/\langle S(\vec{Q})\rangle^2$, which equals one in the ordered phase and zero in the disordered phase.

In contrast to frustrated magnetism, at $\tilde{\mu}=0$ which corresponds to zero longitudinal magnetic fields, the $Z_2$ symmetry cannot be preserved when coupled to the cavity-QED. Consequently, the Rydberg state occupation density $\rho_b=\langle\sum_in_i^{(b)}\rangle/N$ is slightly larger than 1/2 in the SRC phases at small $g$ (see inset of Fig.~\ref{fig4}(a)). Meanwhile, as shown in Fig.~\ref{fig4}(a), neither the structure factor nor the Binder cumulant exhibits any discontinuity along the variation of $g$ for different system sizes up to $L=36$. This indicates that the QPT from the SRC phase to the SR phase is continuous at $\tilde{\mu}=0$. To determine the critical exponents, we performed finite-size scaling analysis via the data collapse of the Binder cumulant \cite{dqcp_zhang}. As illustrated in Fig.~\ref{fig4}(b), after rescaling the atom-photon coupling as $L^{1/{\nu}}(g/g_c-1)$, the Binder cumulants for different system sizes collapse into a single curve, with a fitting critical point of $g_c=2.440\pm0.00072$. Additionally, the extracted critical exponent $1/\nu=1.50\pm0.09$ closely matches that of the 3D XY model at the finite-temperature phase transition $\nu=0.662(7)$ \cite{gottlobCriticalBehaviour3D1993}. This suggests that the second-order phase transition from the SRC phase to the SR phase likely belongs to the 3D XY universality class \cite{obd_moessner,obd_moessner2,ss_prokof,ss_lars}.

Similar to the Ginzburg-Landau theory of the XXZ model \cite{ss_melko}, the stability of the SRC is expected to be determined by the competition between the threefold clock term $M |\psi|^3\cos{(3\theta)}$ and the sixfold clock term $|\psi|^6\cos{(6\theta})$, where $\psi=|\psi|e^{i\theta}$ is the order parameter. In the TFIM, the absence of a threefold clock term allows the OBD phase (or clock phase) to emerge. On the other hand, for the XXZ model away from the $Z_2$ symmetry line, the nonzero uniform magnetization $M$ makes the threefold term relevant, leading to the emergence of the SS phases. However, the nature of the first-order QPT at the $Z_2$ symmetry line remains debated \cite{ss_prokof,ss_lars,XXZ-energy-gap}, with a possible tiny spontaneous breaking $M$ potentially causing serious finite-size effects. Fortunately, here $M$ is expected to be modified to $\rho_t-1/2$ and remains positive at $\tilde{\mu}=0$ due to the small photon density. Consequently, the system flows into the SRC II phase with a higher $\rho_b>0.5$. This analysis is supported by the histogram of $\psi$. At small $g$ (Fig.~\ref{fig4}(c)), three clear peaks indicate $(-,+,+)$, confirming the system’s presence in the SRC II phase. As the system approaches the critical point (Fig.~\ref{fig4}(d)), an irregular circle emerges, hinting at a possible emergent $U(1)$ symmetry (more quantitative discussion shown in SM)\cite{sm}. Given that $\rho_b$ attempts to approach $1/2$ when $g\ge g_c$ (see inset of Fig.~\ref{fig4}(a)), the revival of $Z_2$ symmetry might render the threefold clock term irrelevant at the critical point. Considering that the numerical simulation is performed in a finite system, a possible weakly first-order QPT, rather than belonging to the 3D XY universality class, cannot be entirely ruled out in larger system sizes.

\begin{figure}[t]
	\centering
	\includegraphics[width=0.99\linewidth]{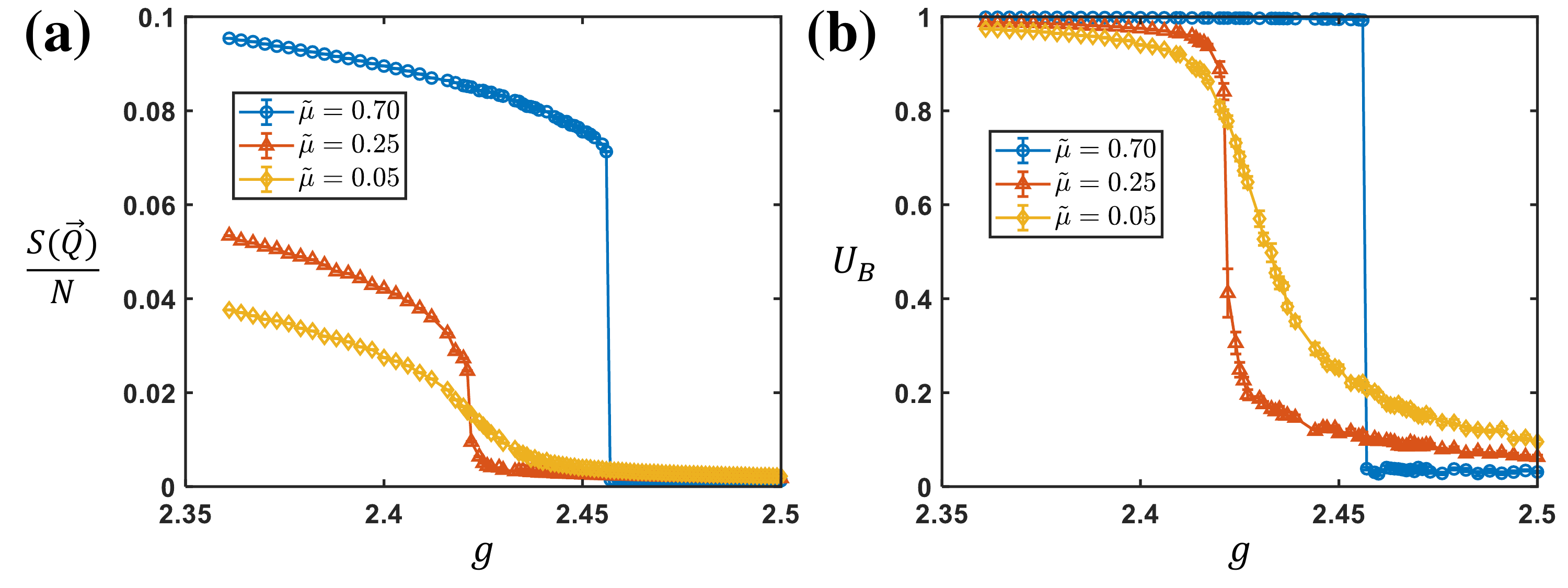}
	\caption{(a) Structure factor $S(\vec{Q})/N$ and (b) its Binder cumulant $U_B$ calculated by QMC simulation across the SRC-SR transition away from the $Z_2$ symmetry line for $L=24$.
}\label{fig3}
\end{figure}

Away from the $Z_2$ symmetry line, the QPT between SRC phases and SR phase changes from continuous to first-order, which is directly reflected by the discontinuity of both the structure factor and the Binder cumulant, as shown in Fig.~\ref{fig3}. The first-order QPT begins with two triple points $(\tilde{\mu},g)\approx(1.00,2.461)$ and $(-0.95,2.850)$. As the magnitude of $\tilde{\mu}$ close to the 3D XY point, the discontinuities of the structure factor become systematically smaller. In Fig.\ref{fig3}, we can observe that the first-order phase transition terminates at the tri-critical point $(\tilde{\mu},g)\approx(0.25,2.421)$ in the case of QPT between SRC II and SR phase under finite system size $L=24$. However, we still cannot rule out the possibility that the appearance of the tri-critical point \cite{ss_zhang} is due to the finite size effect, and they may also emerge or merge with the 3D XY critical point in the thermodynamic limit. Discussion about the possible tri-criticality can be found in SM \cite{sm}.

\begin{figure}[t]
	\centering
	\includegraphics[width=0.99\linewidth]{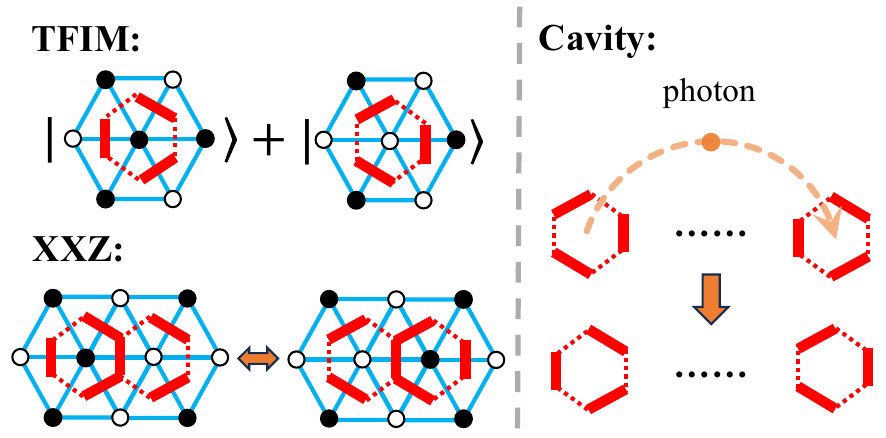} 
	\caption{Dimer representation of ring exchange process in the TFIM, XXZ model, and Rydberg atom array in a cavity.
 }\label{fig5}
\end{figure}

\textit{Quantum Dimer Model}.-- The ground state of frustrated magnetism can typically be described using dimer language. As shown in Fig.~\ref{fig5}, bonds with the same states can be mapped to dimers, allowing the Rydberg occupation configuration in the triangular lattice to be mapped to the dimer configuration in the dual honeycomb lattice \cite{obd_moessner,obd_moessner2}. Consequently, the local constraint in each triangle is transferred to a single dimer linking to the dual site, and all possible degenerate ground states can be classified into different topological sectors based on winding numbers \cite{lgt_zhang1} (More discussions in SM \cite{sm}). Compared to the solid phase, the disordered state with high degeneracy at half-filling is expected to be unstable under the perturbation of atom-photon coupling. Considering an atom in the Rydberg state surrounded by an alternatively occupied hexagon depicted as $\hexa$, it can return to the state $\hexb$ and emit a photon without violating the local constraint. In dimer language, this corresponds to the ring exchange in the hexagon $|\dima\rangle\langle\dimb|$. Meanwhile, as demonstrated in Fig.~\ref{fig5}, another ring exchange can occur by absorbing a photon from a distant location. Thus, the cavity-mediated long-range ring exchange between two hexagons is nonlocal, similar to the double-hexagon ring-exchange process in the XXZ model \cite{ss_damle05}. We can argue that such nonlocal ring exchange may significantly lower the energy of the SRC phase, akin to the effect of XY interactions \cite{XXZ-energy-gap}.

\textit{Conclusion and Discussion.--} Our QMC simulations of Rydberg atomic triangular arrays coupled with cavity-QED reveal the emergence of novel SRC phases and rich QPT phenomena, driven by the competition between sixfold and threefold clock terms. In real experiments, Rydberg atom interactions extend beyond nearest neighbors, but both experimental \cite{Rydberg_nature2} and numerical \cite{guoOrderDisorderEmergent2023} studies indicate that long-range interactions only slightly alter the OBD phase region which is also discussed in SM \cite{sm}. Certainly, the impact of long-range interactions on SRC phases, particularly their interplay with emergent U(1) lattice gauge theory \cite{lgt_zhang1,lgt_zhang2}, warrants further investigation. Especially, the recent advance of the Rydberg dressing atom in the optical lattice could make the system to be more larger and meanwhile get rid of the long-range interaction tail's influence \cite{ebhm}.

Our simulations assume a perfect, non-leaking cavity. However, cavity imperfections in real experiments could significantly affect the OBD mechanism, the influences of the cavity leakage and inhomogeneity are discussed in SM \cite{sm}. Other frustrated geometries coupled with cavity-QED, such as the Kagome lattice \cite{Rydberg_kagome,cavity_kagome} and spin ice \cite{Rydberg_spinice}, may also present intriguing possibilities for disorder-by-disorder phenomena and emergent QED physics. In summary, we believe that cavity-mediated infinite long-range interactions can profoundly alter our understanding of emergent physics in both frustrated systems and many-body quantum optics.

\section*{acknowledgments}
We would like to thank Changle Liu for many helpful discussions and especially for sharing the data of Ref.\cite{liuIntrinsicQuantumIsing2020}. X.-F. Z. acknowledges funding from the National Science Foundation of China under Grants  No.12274046 and No.12547101, and Xiaomi Foundation / Xiaomi Young Talents Program.

\bibliography{ref}

\section{Supplemental Material}
\subsection{Variational Approach}
The ground state phase diagram can also be calculated by using the variational approach, and the corresponding variational wave function can be explicitly written as
\begin{equation}
\ket{\lambda,\theta_i}=e^{-\frac{\lambda\sqrt{N}}{2}a^{\dagger}}\otimes\prod_i{\left[\cos{\left(\frac{\theta_i}{2}\right)}b^\dagger_i+\sin{\left(\frac{\theta_i}{2}\right)}\right]}\ket{0},
\end{equation}
where $i=A, B, C$ labels the three sublattices, $\lambda$, and $\theta_i$ are the variational parameters for the quantum photonic field and Rydberg atom states. Then, the variational energy per site can be calculated and equals
\begin{align}
E &= \frac{\bra{\lambda,\theta}H\ket{\lambda,\theta}}{N}  
\notag
\\&= -\frac{g\lambda}{6}\left(\sin{\theta_A}+\sin{\theta_B}+\sin{\theta_C}\right)
-\frac{\lambda^2 \mu}{4} 
\notag
\\&-\frac{1}{6}(\cos{\theta_A}+\cos{\theta_B}+\cos{\theta_C}+3)\left(\mu+\Delta\right)  
\notag  
\\&+\frac{V}{4}(\cos{\theta_A}\cos{\theta_B}+\cos{\theta_B}\cos{\theta_C}+\cos{\theta_C}\cos{\theta_A}
\notag  
\\&+2\cos{\theta_A}+2\cos{\theta_B}+2\cos{\theta_C}+3).
\label{eq9}
\end{align}
From the expression, we find out that the region of the variational parameters can be limited to $\lambda\ge0$ and $0\le\theta_A\le\theta_B\le\theta_C\le\pi$ so that the minimisation of the variational energy can be simplified. The different phases can be characterised by the variational values listed in Table \ref{table1}. 
\begin{table}[h]
	\centering  
	\begin{ruledtabular}
		\begin{tabular}{cccccccc}
			Parameters & SRC I & SRC II & SRC 0 & solid(1/3) & solid(2/3)  & SR\\
			\colrule
			$\lambda$  & $\ne0$       & $\ne0$       & $\ne0$       & $0$   & $0$   & $\ne0$     \\
			$\theta_A$ & $\theta_1<\frac{\pi}{2}$ & $\theta_1<\frac{\pi}{2}$ & $\theta_1$ & $0$ & $0$ & $\theta$ \\
			$\theta_B$ & $\theta_3>\frac{\pi}{2}$ & $\theta_1<\frac{\pi}{2}$ & $\theta_2$ & $0$   & $\pi$ & $\theta$ \\
			$\theta_C$ & $\theta_3>\frac{\pi}{2}$ & $\theta_3>\frac{\pi}{2}$ & $\theta_3$ & $\pi$   & $\pi$   & $\theta$ \\
		\end{tabular}
		\caption{The variational values for different phases, where $\theta_1<\theta_2<\theta_3$ in SRC 0 phase. These values are determined by minimizing Eq.(\ref{eq9}). }
		\label{table1}
	\end{ruledtabular}
\end{table}

\begin{figure}[t]
	\centering
	\includegraphics[width=0.9\linewidth]{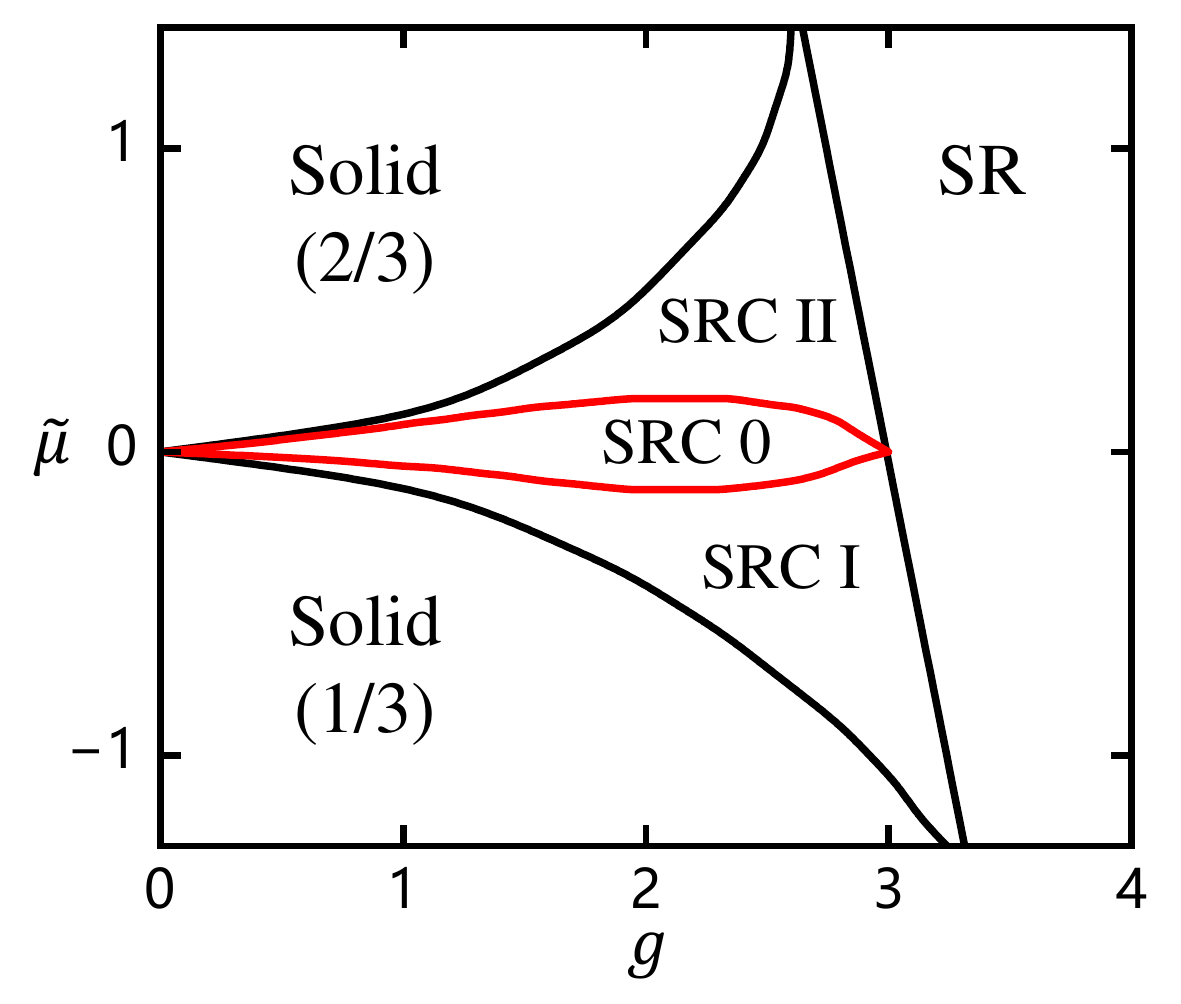}
	\caption{Variational phase diagram for $\Delta=9$ and $V=1$.
	}\label{figs1}
\end{figure}

\begin{figure}[t]
	\centering
	\includegraphics[width=0.99\linewidth]{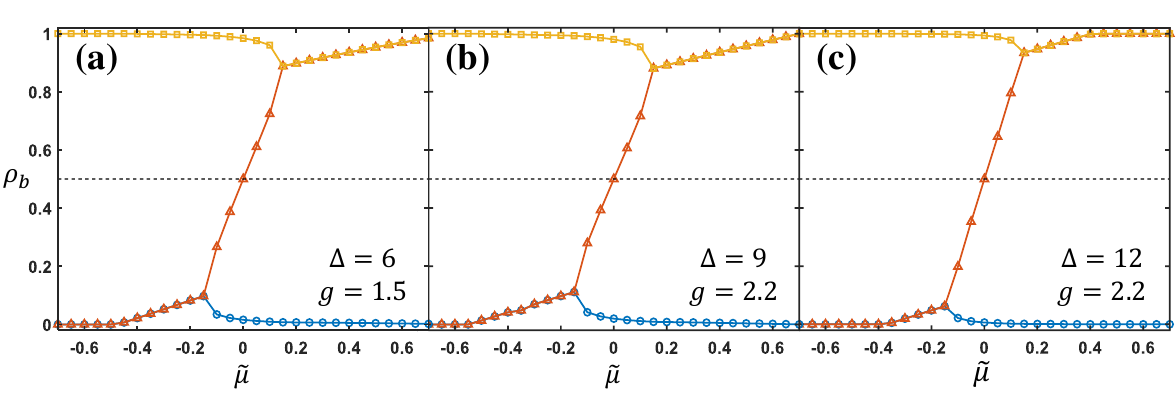}
	\caption{Variational results of the Rydberg state occupation density in three sublattices under different detuning.
	}\label{figs2}
\end{figure}

The variational phase diagram is shown in Fig.\ref{figs1}, and we can find an intermediate phase between SRC phases instead of the first-order QPT. To check its structure, we calculate the Rydberg state occupation density of the three sublattices $\rho_b^i=\cos^2{(\theta_i/2)}$ , which are plotted in Fig.\ref{figs2}. Surprisingly, at $Z_2$ symmetry point, the local magnetisation presents $(+,0,-)$ structure which is the same as the OBD phase. Considering its photon density is nonzero, we name it as SRC 0 phase. The $(+,0,-)$ structure indicates the SRC 0 phase is driven by the sixfold clock term. In the variational approach, the quantum photon field is replaced with a coherent state of light, which is the most classical state of light, so the quantum property of the light is heavily underestimated. Therefore, by comparing with the quantum phase diagram via the QMC method in Fig.1 in the main text, we think the SRC I and II phases can gain more energy from quantum fluctuations so that the OBD phase (or SRC 0 phase) is totally suppressed. Furthermore, in the real experiment, the leaking of the cavity makes the light behave more ``classical", so the intermediate SRC 0 phase may emerge in the frustrated open system coupled with the cavity.

\subsection{Strong Coupling Expansion Method}
The phase transition from the solid phases to the SRC phases is continuous and induced by exciting the polariton on the honeycomb lattice backbone. Considering the QPT happens at a small $g$, it is possible to utilize the strong coupling expansion (SCE) method by taking the atom-light coupling interaction as a perturbative term. 

Therefore, the corresponding critical lines can be calculated. For the phase transition from solid (1/3) to SRC I phase, we can calculate the second-order perturbation energy of the solid (1/3) phase as $E_{1/3}=-N(\tilde{\mu}/3+V)+g^2/3\Delta$. When entering the SRC I phase, we consider excitation of a Rydberg state based on the 1/3-filled background and obtain the second-order perturbation energy: $E_{1/3}'=-(\tilde{\mu}+3V)(N/3-1)+3V-2g^2/(3\Delta-9V)-g^2/3\Delta$. At the phase boundary,  $E_{1/3}$ and $E_{1/3}'$ should be equal, so we can determine the phase boundary:
\begin{equation}
\tilde{\mu}_{c1}=-\frac{2g^2}{3(\Delta-3V)}.
\label{eq4}
\end{equation}

\begin{figure}[b]
	\centering
	\includegraphics[width=0.75\linewidth,]{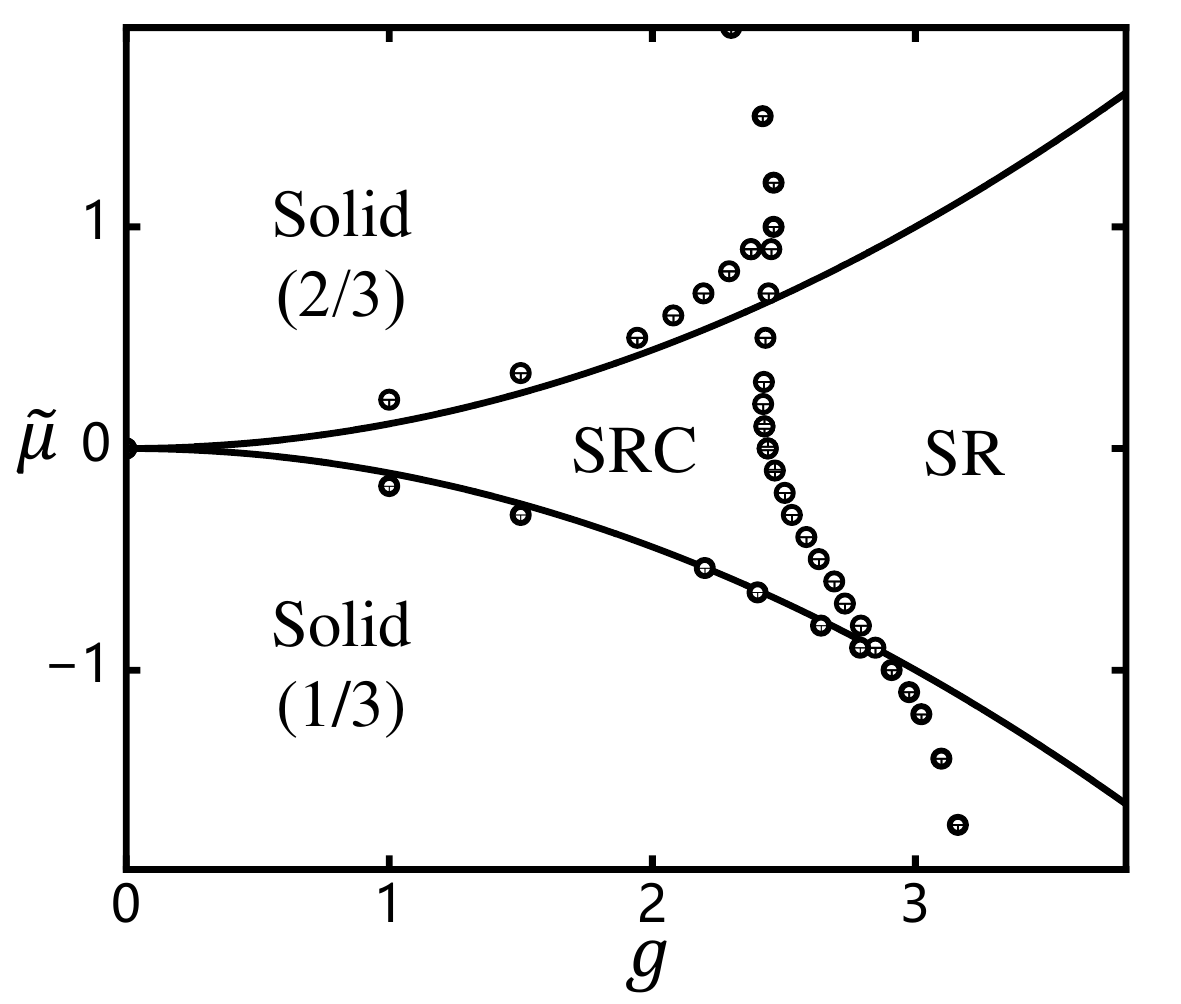}
	\caption{Phase boundaries determined by SCE (solid line) and QMC (circle) with same parameters as Fig.1 in the main text.
	}\label{figs3}
\end{figure}
Similarly, for the phase transition from the solid (2/3) phase to the SRC II phase, the second-order perturbation energies are: $E_{2/3}=-2N(\tilde{\mu}/3+V)+NV+2g^2/(3\Delta-9V)$ and $E_{2/3}'=-(\tilde{\mu}+3V)(2N/3-1)+(N-3)V-4g^2/(3\Delta-9V)$, which correspond to the 2/3-filled state and its hole excitation. Thus, we can determine the phase boundary
\begin{equation}
\tilde{\mu}_{c2}=\frac{2g^2}{3(\Delta-3V)}.
\label{eq5}
\end{equation}

As demonstrated in Fig.\ref{figs3}, the phase boundary given by SCE is close to the QMC results, especially the lower critical line $\tilde{\mu}_{c1}$. From Fig.2 (b) in the main text, we can find that the photon density around the upper critical line $ \tilde{\mu}_{c2}$ is larger than $\tilde{\mu}_{c1}$, so the large deviation of the upper critical line may result from the high order perturbation's non-negligible contribution.

\subsection{Influence of the Detuning}

\begin{figure}[b]
	\centering
	\includegraphics[width=0.99\linewidth,]{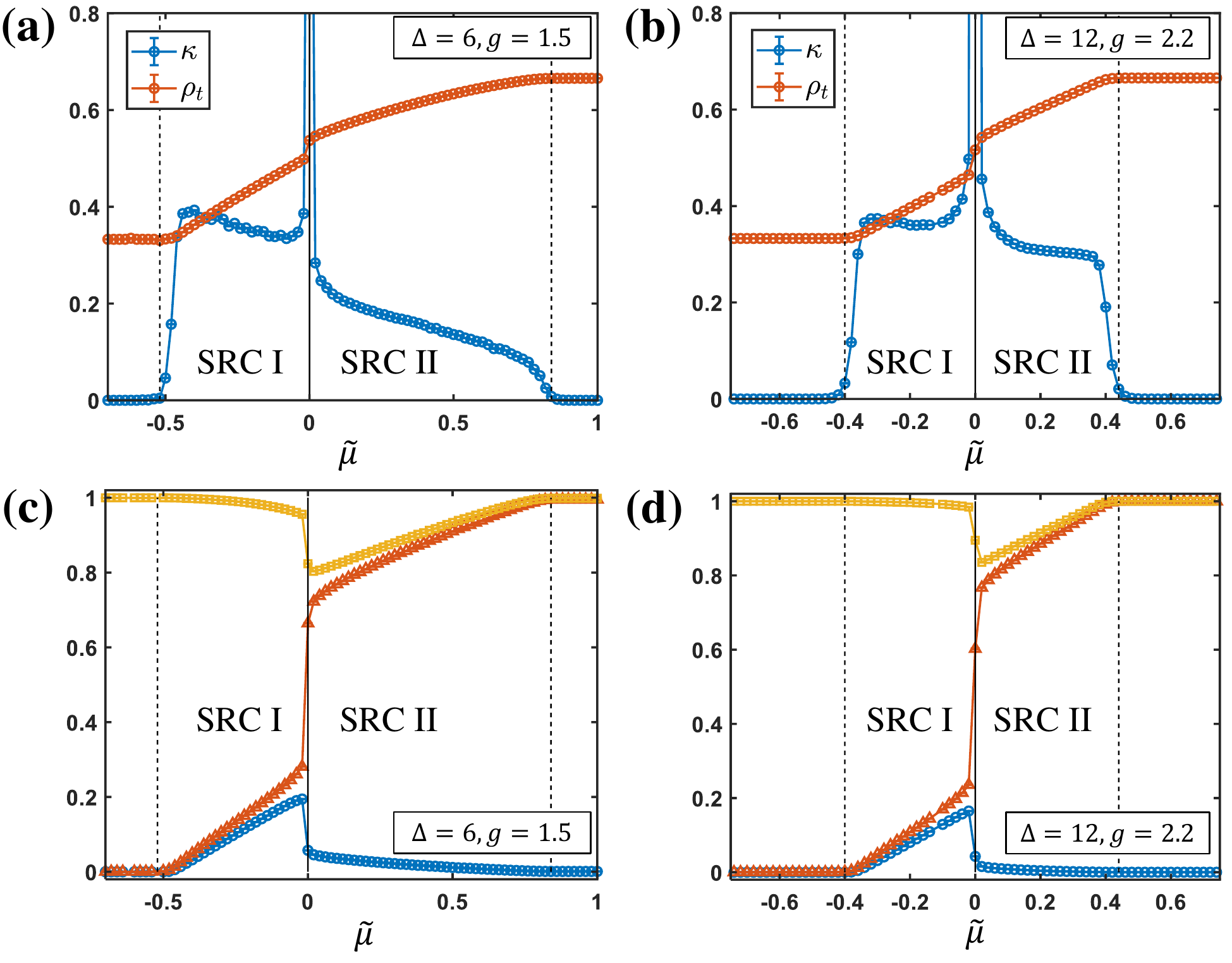}
	\caption{(a,b) Compressibility $\kappa$, the average total density $\rho_t$, and (c,d) Rydberg state occupation density in three sublattices calculated by QMC simulation at $L=24$ with different detuning.
	}\label{figs4}
\end{figure}

The experiment in the canonical ensemble can be transformed to the following Hamiltonian in the grand canonical ensemble \cite{anQuantumPhaseTransition2022}:
\begin{eqnarray}
\nonumber
    H&=&\frac{g}{\sqrt{N}}\sum^{N}_{i=1}{\left(b^{\dagger}_ia+a^{\dagger}b_i\right)}+\sum^{N}_{<ij>}V_{ij}{n_i^{(b)}n_j^{(b)}}+\omega_c n^{(a)}+\\
    &&\epsilon\sum^{N}_{i=1} n_i^{(b)}-\mu_tN_t,
    \label{eq1}
\end{eqnarray}
where $\omega_c$ and $\epsilon$ are the cavity and atom transition frequencies with the detuning defined by $\Delta=\omega_c-\epsilon$. Then, after comparing with Eq.(1) in the main text, we can immediately find out $\mu=\mu_t-\omega_c$ and $\mu_b=\mu_t-\epsilon=\mu_t-(\omega_c-\Delta)=\mu+\Delta$. Meanwhile, to reflect the ${Z}_2$ (or spin-up-down) symmetry, we introduce $\tilde{\mu}=\mu_b-3V$, so we can get the relation $\mu=\tilde{\mu}+3V-\Delta$. In the numerical simulation, the values of $V=1$ and $\Delta=9$ are fixed, so the $\tilde{\mu}-g$ phase diagram can be directly transformed to the $\mu-g$ phase diagram. Therefore, although $\mu$ is related to the frequency of the photon $\omega_c$, adjusting $\omega_c$ and $\epsilon$ in the experiment is equivalent to tuning the value of $\Delta$. Then, it would be necessary to discuss the influence of the detuning.

According to the Ginzburg-Landau theory discussed in the main text, the photon number density of the system is likely to influence the energy gap of the first-order phase transition. To ensure that the first-order phase transition shown in Fig.~2 in the main text remains stable under different detunings, we calculated the compressibility $\kappa$ and the total particle number density $\rho_t$ for different detunings. The results presented in Fig.~\ref{figs4}(a,b) demonstrate that the signal of the first-order phase transition remains robust under different detunings.

We can also examine the Rydberg state occupation density in three sublattices at different detunings. As demonstrated in Fig.~\ref{figs4}(c,d), the local magnetization in the SRC I and II phases shows a clear feature of sign structure $(-,-,+)$ and $(-,+,+)$, respectively. Compared to the variational results, the densities at the backbone two sublattices are different. This discrepancy may be attributed to the possible SRC 0 phase due to the finite temperature effects.

\begin{figure}[h]
    \centering
    \includegraphics[width=0.75\linewidth]{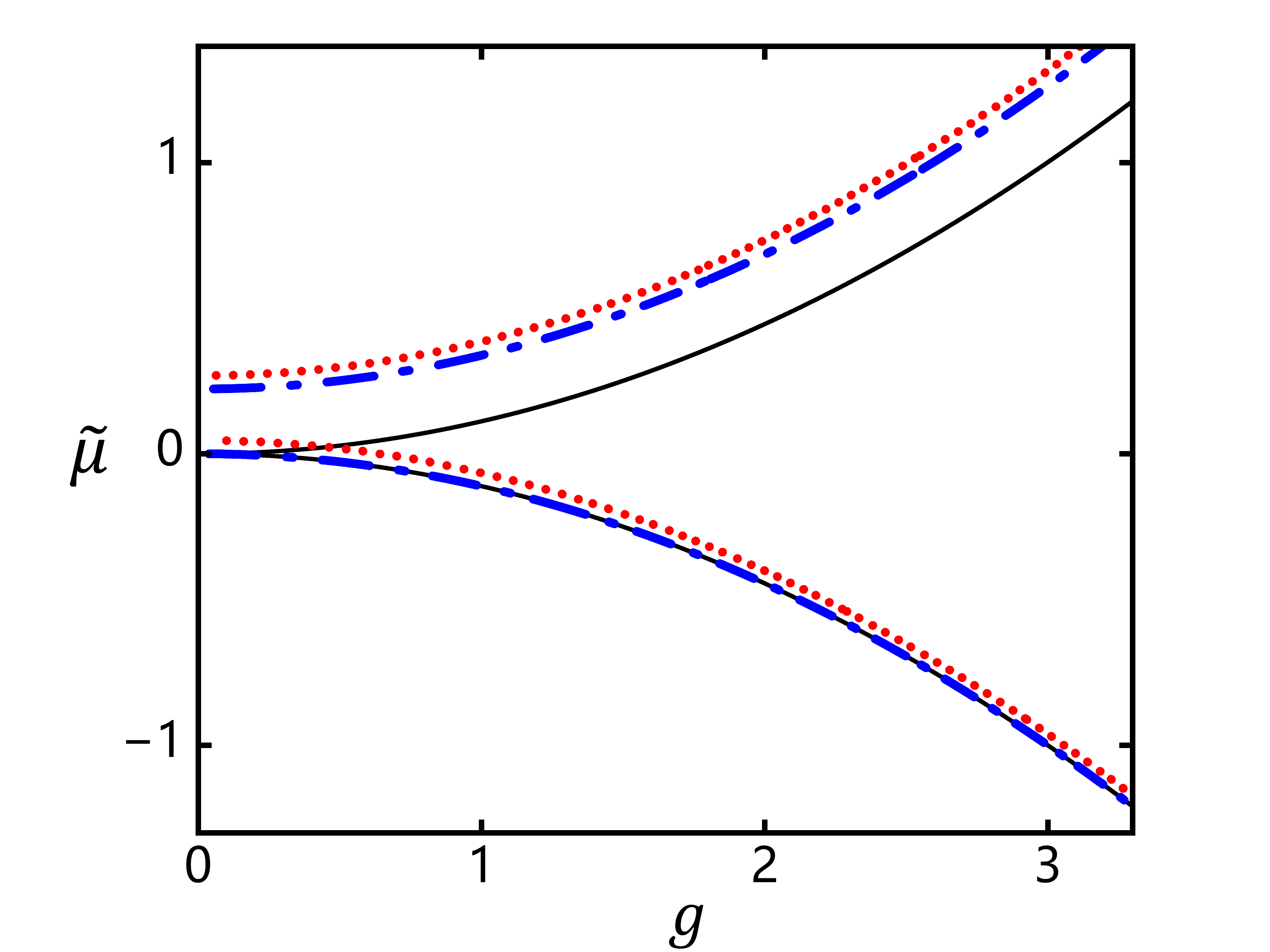}
    \caption{Phase boundaries between SRC and solid phase, determined by the SCE with $\Delta=9$ and Rydberg repulsive interaction truncated up to nearest-neighbor (black solid line), next nearest-neighbor (blue dot dashed line), and next next nearest-neighbor sites (red dotted line).}
    \label{figs5}
\end{figure}

\begin{figure}[h]
    \centering
    \includegraphics[width=0.75\linewidth]{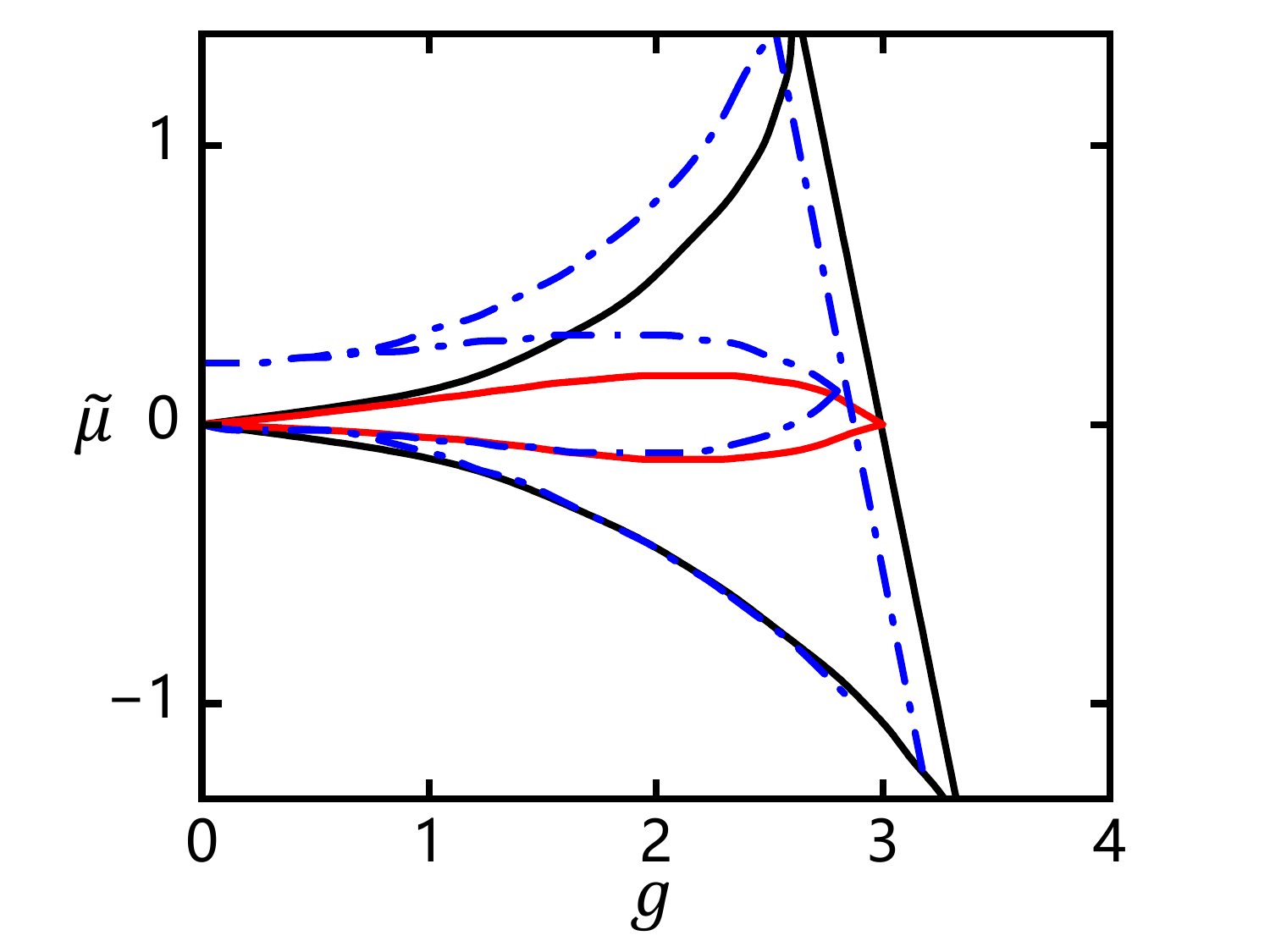}
    \caption{Phase diagram calculated via variational approach at $\Delta=9$. The blue dot dashed line takes into account the NN interaction, in comparison with the solid lines shown in Fig.\ref{figs1}.}
    \label{figs6}
\end{figure}

\subsection{long-range interaction}
Since the VdW interaction between Rydberg atoms is actually a long-range interaction that decays with $r^{-6}$, in this section, we discuss the effect of further truncation distances on the phase diagram. In the main text, we only consider the nearest neighbor interaction, but here we extend it to the 3rd nearest neighbor interaction with the Hamiltonian below:
\begin{equation}
    H=\frac{g}{\sqrt{N}}\sum^{N}_{i=1}{\left(b^{\dagger}_ia+a^{\dagger}b_i\right)}+\sum^2_{k=0}\sum^{N}_{ij}V_k{n_i^{(b)}n_j^{(b)}}-\mu n^{(a)}-\mu_b\sum^{N}_{i=1} n_i^{(b)},
    \label{eqs4}
\end{equation}
where $V_0=1$, $V_1=1/27$, $V_2=1/64$ corresponding to the nearest neighbor(N), next nearest neighbor(NN) and 3rd nearest  neighbor(NNN) interactions. With the help of the SCE, we can obtain the analytic expression of the phase boundaries between SRC and solid phases, which are presented in Table.\ref{SCEtable}. As shown in Fig.\ref{figs5}, the long-range interaction can only impose a tiny shift. Meanwhile, the results of the variational method (Fig.\ref{figs6}) indicate that the long-range interactions that decay with $r^{-6}$ do not destroy any phases, but only slightly deform the whole phase diagram. 

\begin{figure}[t]
    \centering
    \includegraphics[width=0.99\linewidth]{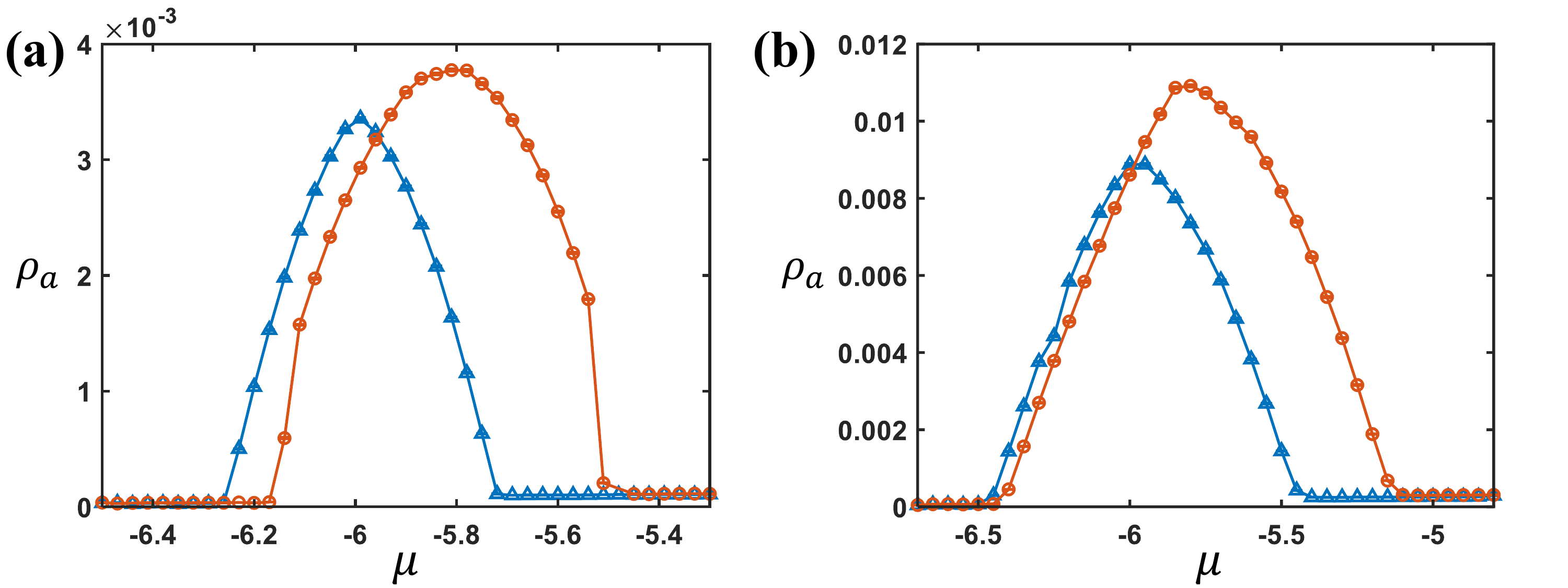}
    \caption{The photon density $\rho_a$ vs chemical potential $\mu$ calculated by QMC simulation at (a) $g=1.50$ and (b) $g=2.00$ with $\Delta=9$, $L=24$, and the Rydberg repulsive interaction truncated up to nearest-neighbor (blue triangle) and 3rd nearest-neighbor sites (red circle).}
    \label{figs7}
\end{figure}

\begin{figure}[h]
    \centering
    \includegraphics[width=0.75\linewidth]{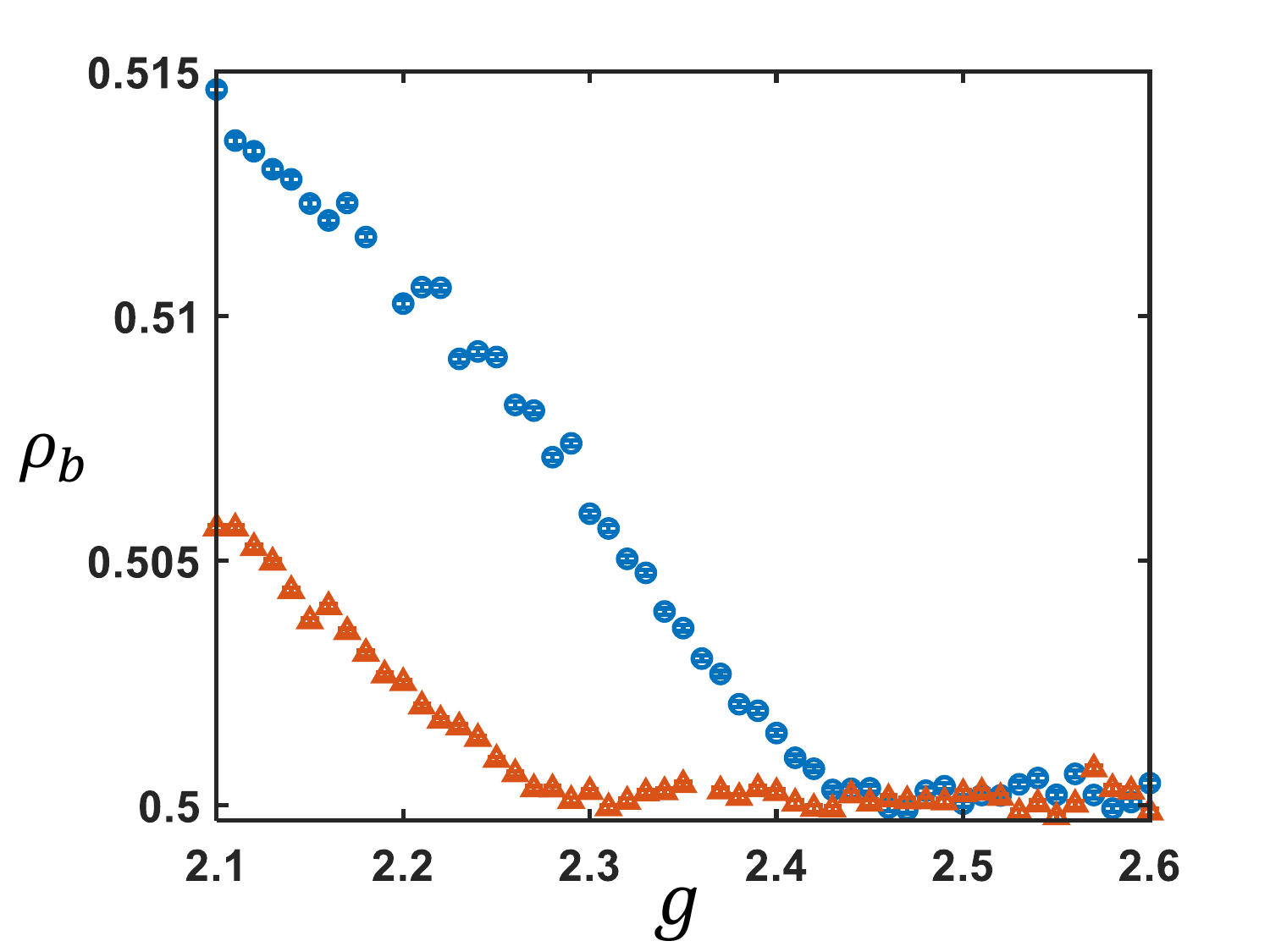}
    \caption{The Rydberg density $\rho_b$ vs $g$ calculated by QMC simulation at $Z_2$ symmetry line with $\Delta=9$, $L=24$, and the Rydberg repulsive interaction truncated up to nearest-neighbor (blue triangle) and 3rd nearest-neighbor sites (red circle).}
    \label{figs8}
\end{figure}

The shift of the phase boundaries can be verified by the simulation of QMC, and is clearly reflected in both Fig.\ref{figs7} and Fig.\ref{figs8}. Meanwhile, along the $Z_2$ symmetry line (Fig.\ref{figs8}), we can still observe that the Rydberg atom density $\rho_b>0.5$, indicating that the ground state is still governed by the $M|\psi|^3\cos{(3\theta)}$ term. However, maybe due to the possible glassiness effect introduced by the long-range interaction, it becomes easier to destroy the crystal order, and the critical value $g_c$ decreases to be smaller.

\begin{table*}
    \centering
    \begin{tabular}{|c|c|c|}\hline
         &  N($V_0$)+NN($V_1$)& N($V_0$)+NN($V_1$)+NNN($V_2$)\\\hline
         Solid(1/3)-SRC I&  $\tilde{\mu}=-\frac{2g^2}{3}\frac{1}{\Delta-3V_0}$
 & $\tilde{\mu}=-\frac{2g^2}{3}\frac{1}{\Delta-3V_0-3V_2}+3V_2$\\\hline
         Solid(2/3)-SRC II&  $\tilde{\mu}=\frac{2g^2}{3}\frac{1}{\Delta-3V_0-6V_1}+6V_1$& $\tilde{\mu}=-\frac{2g^2}{3}\frac{1}{\Delta-3V_0-6V_1-3V_2}+6V_1+3V_2$\\ \hline
    \end{tabular}
    \caption{Analytical expression of the phase boundaries solved by SCE for different truncation distances.}
    \label{SCEtable}
\end{table*}

\subsection{cavity leaking}
In the experiment, the cavity-photon leakage is usually on the same order of magnitude as $\tilde{g}=g/\sqrt{N}$\cite{ligang}, which generally gives rise to non-negligible decoherence effects. To investigate whether this level of photon loss affects the stability of the SRC phases, we conduct quantum-trajectory-based simulations using QuTiP \cite{qutip5}, an open-source Python package. Limited by computational power, we considered a $3\times3$ system with a photon cutoff of $8$. 

\begin{figure}
    \centering
    \includegraphics[width=0.99\linewidth]{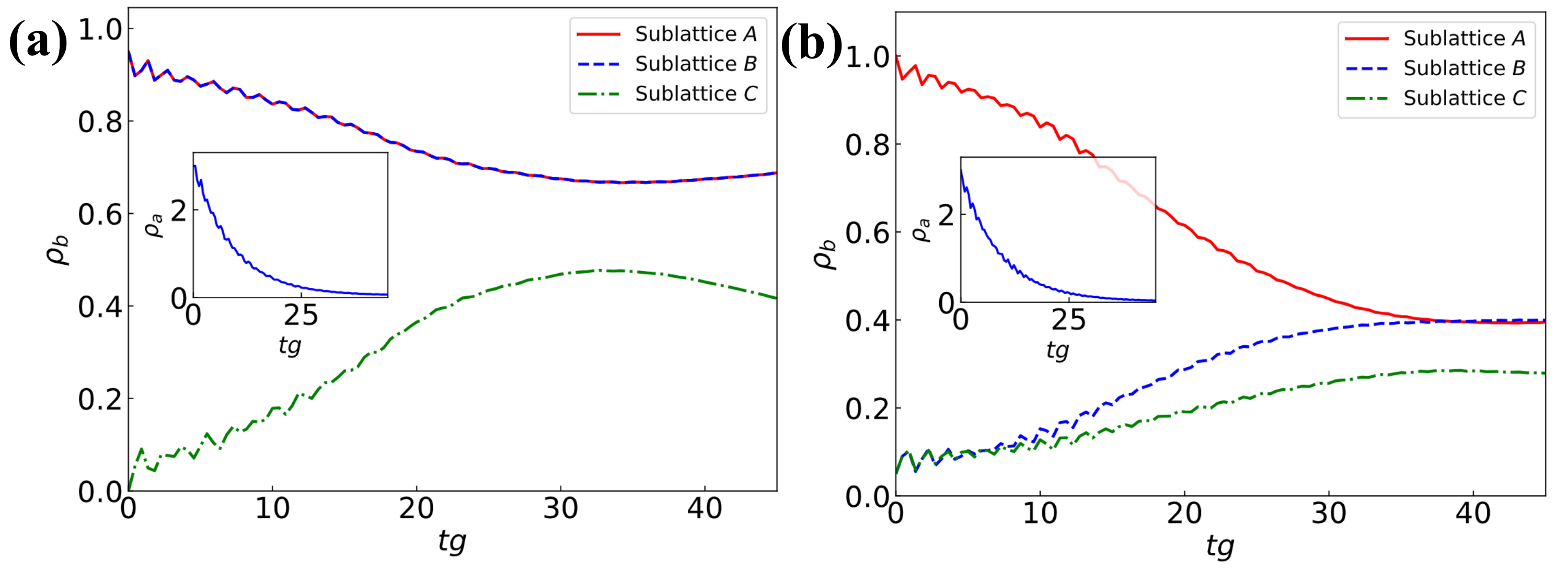}
    \caption{The evolution of Rydberg atom density $\rho_b$ in three sublattices and the photon density $\rho_a$ (inset) with only cavity leakage rate $\tilde{\kappa}=\tilde{g}/3$. The initial state: (a) SRC II phase at $\tilde{\mu}=0.3$, $g=1.8$, (b) SRC I phase at $\tilde{\mu}=-0.2$, $g=1.8$.}
    \label{figs9}
\end{figure}

In Fig.\ref{figs9} we present the case where only cavity-photon leakage with leaking rate $\tilde{\kappa}=\tilde{g}/3$ is considered. Fig.\ref{figs9}(a) shows that, within the range of our numerical simulation, the three-fold clock order of the SRC-II phase is not destroyed by cavity-photon decay. In contrast, for the SRC-I phase (Fig.\ref{figs9}(b)), we observe that the three-fold clock order survives only for the first several Rabi periods. After that, the Rydberg atom density in one sublattice (B) begins increasing faster than the other sublattice (C). It indicates the six-fold clock order begins to take effect, which may result from the decoherence of the photon induced by photon leakage and is consistent with the prediction of the variational approach in Fig.\ref{figs1}. 

\begin{figure}
    \centering
    \includegraphics[width=0.99\linewidth]{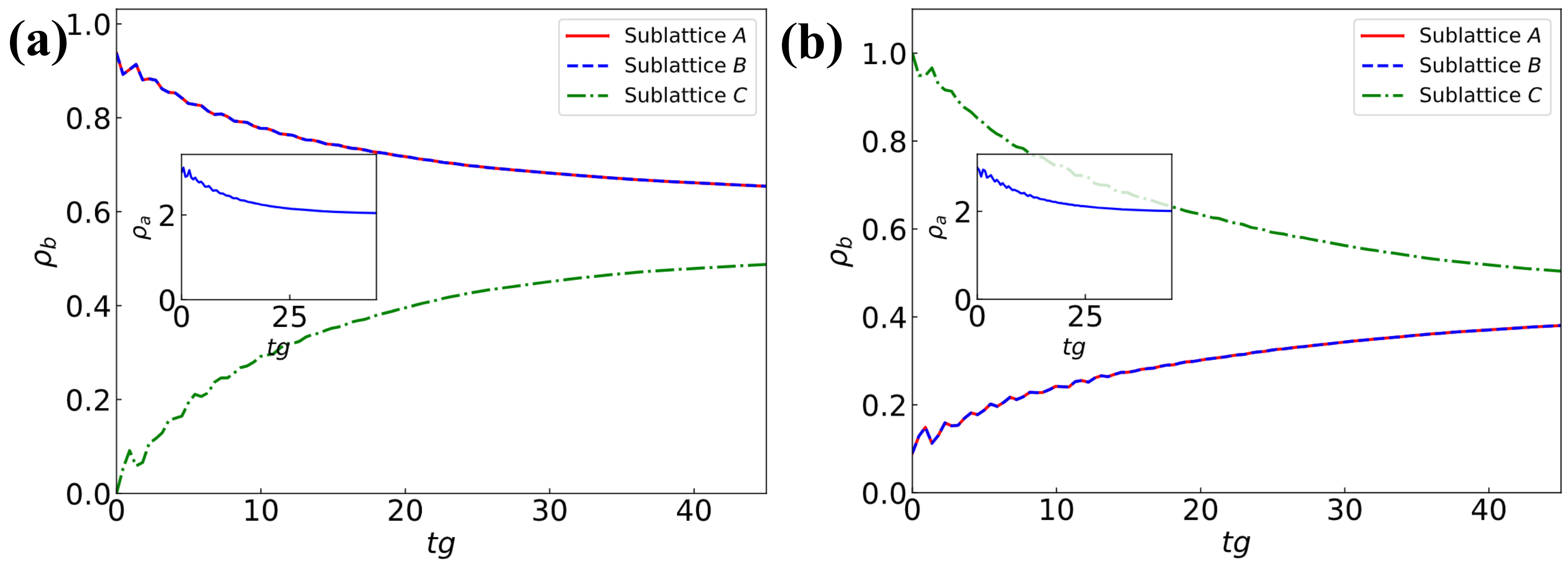}
    \caption{The evolution of Rydberg atom density $\rho_b$ in three sublattices and the photon density $\rho_a$ (inset) with cavity leakage rate $\tilde{\kappa}=\tilde{g}/3$ and pumping rate $\zeta=\tilde{\kappa}$. The initial state: (a) SRC II phase at $\tilde{\mu}=0.3$, $g=1.8$, (b) SRC I phase at $\tilde{\mu}=-0.3$, $g=1.8$.}
    \label{figs10}
\end{figure}

In the experiment, in order to suppress the strong decoherence caused by the cavity leakage, photons are usually pumped into the cavity at the same time. Here, we set the pumping rate $\zeta=\tilde{\kappa}=\tilde{g}/3$ to ensure the photon density will not decay to $0$. As the result shown in Fig.\ref{figs10}, the three-fold clock order always survives both in SRC I and SRC II phases as the photon density $\rho_a$ keeps a finite value during the evolution. In comparison with the no-pumping case, we can find that the coherence of the photon plays a key role in the stability of the SRC I phase.

\begin{figure}[b]
    \centering
    \includegraphics[width=0.99\linewidth]{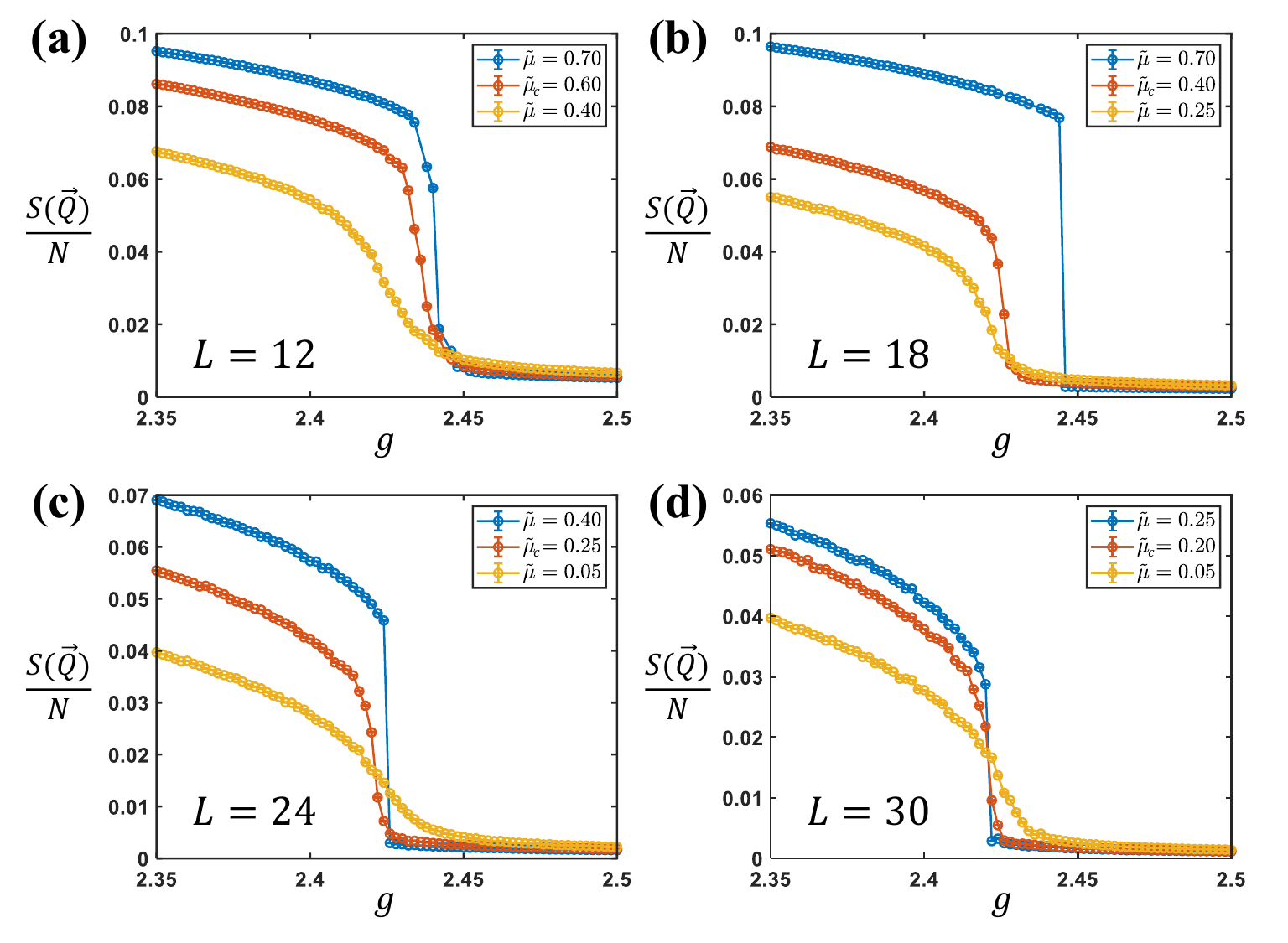}
    \caption{The behavior of the structure factor $S(\vec{Q})/N$ near the phase boundary under different system sizes. The blue and yellow lines demonstrate the first and second order QPT between SRC II and SR phase in the finite-size system. The red lines correspond the finite-size tri-critical points, which equal $\tilde{\mu}=0.60, 0.40, 0.25, 0.20$ for $L=12,18,24,30$.}
    \label{figs11}
\end{figure}

\subsection{tri-critical points}
\begin{figure}[h]
    \centering
    \includegraphics[width=0.75\linewidth]{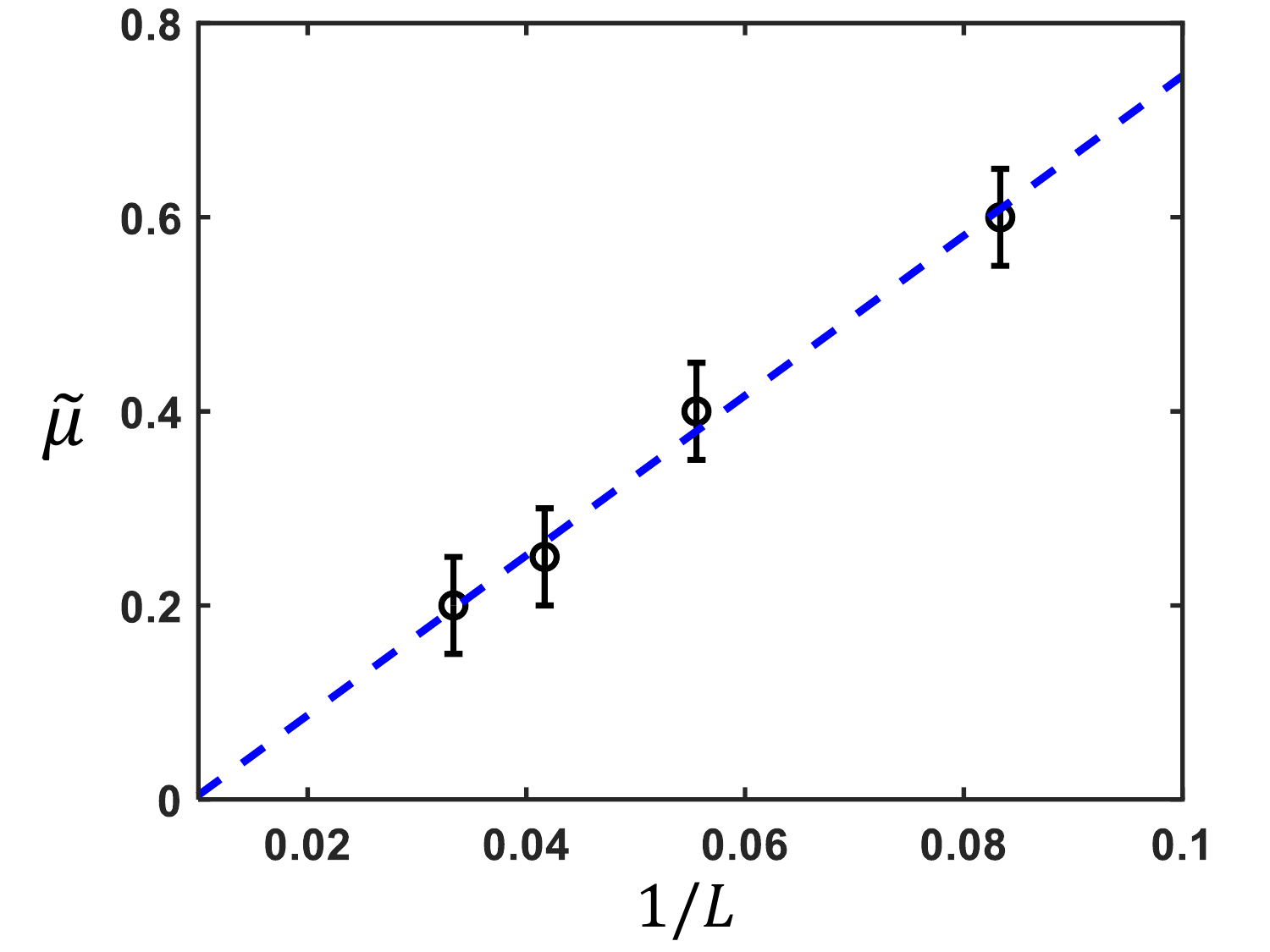}
    \caption{Linear fitting of tri-critical points under finite sizes ($L=12,18,24,30$).}
    \label{figs12}
\end{figure}

In the main text, we show that there are tri-critical points in the QPT between the SRC and SR phases under finite system size $L=24$. Here, we carry out more simulations at different system sizes to figure out whether the tri-critical points will merge together in the thermodynamic limit. According to the QMC simulations in Fig.\ref{figs11}, as the size increases, the finite-size tri-critical points gradually approach the 3D XY point. By performing a linear fitting on these $\tilde{\mu}_c$ values for the inverse of the sizes in Fig.\ref{figs12}, it can be observed that at the thermodynamic limit, the tri-critical points tend to merge with the 3D XY point. It indicates that the appearance of the tri-critical points is very likely to be merely a finite-size effect.

\subsection{anisotropy parameter}
\begin{figure}[h]
    \centering
    \includegraphics[width=0.75\linewidth]{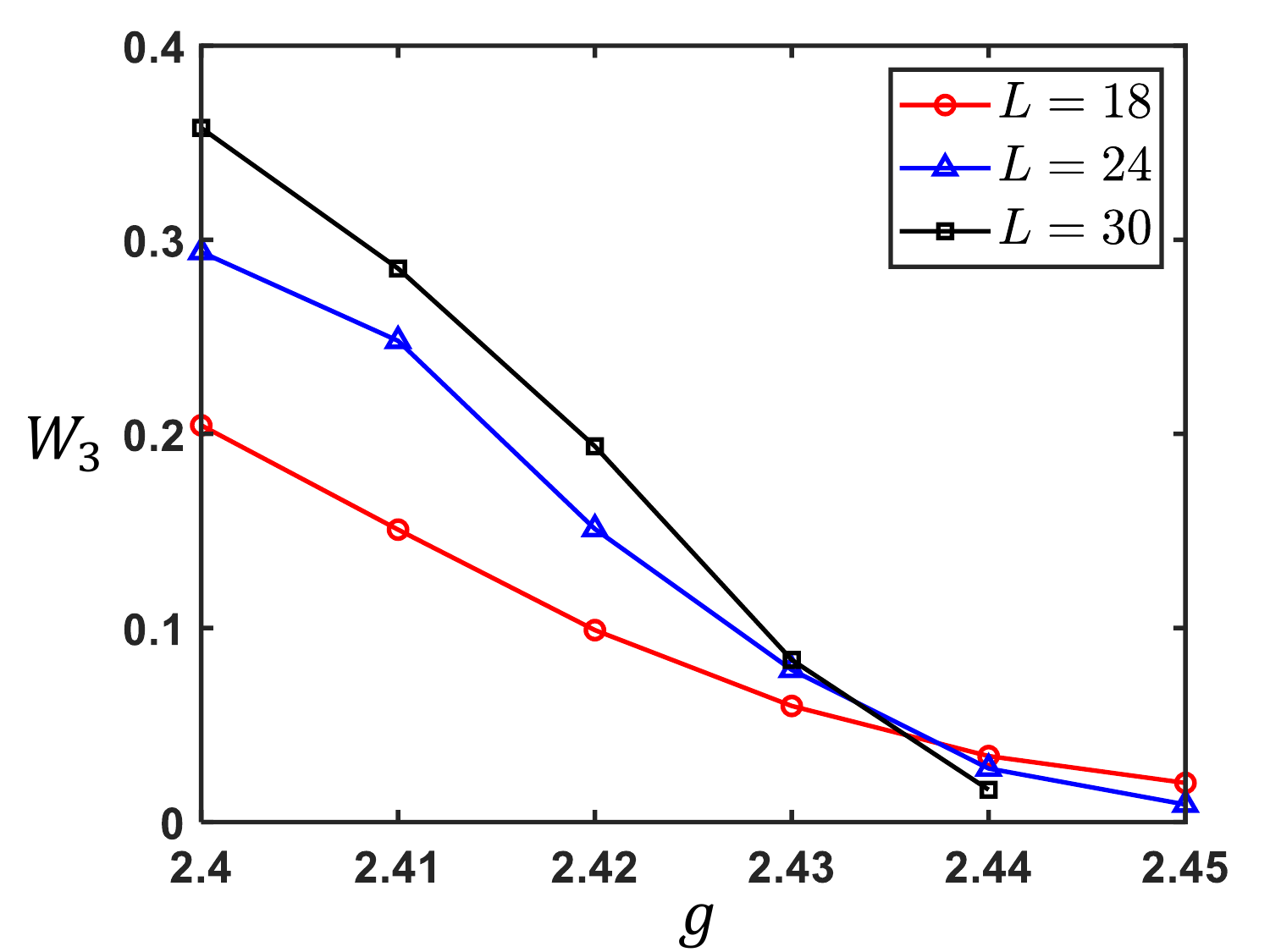}
    \caption{The anisotropy parameter $W_3$ varies with $g$ at different system sizes.}
    \label{figs13}
\end{figure}

Same as other frustrated systems, the irregular circle is usually due to the existence of the marginal or dangerous irrelevant terms. To provide more quantitative evidence, we take into account the anisotropy parameter $W_3=|\cos(3\arg(s(\vec{Q})))|$ \cite{dqcp_zhang}. As shown in Fig.\ref{figs13}, the anisotropy of $s(\vec{Q}) $ near the 3D XY critical point ($g=2.440$) approaches zero while increasing the system size. Therefore, $W_3$ is expected to be zero, and the histogram of $s(\vec{Q})$ can be uniformly $U(1)$ distributed in the thermodynamic limit. Definitely, our data still can not rule out the possibility of the weakly first-order QPT. 

\subsection{Inhomogeneous effect}
How precisely to position the atoms at the antinode of the cavity field is a very challenging technical problem. We have found that this has been achieved quite well in a one-dimensional array \cite{ligang}. They used optical tweezers to trap atoms and simultaneously utilized the blue detuning induced by the cavity light to lock the positions of the atoms more precisely, achieving a variance of the coupling strength between all atoms and the optical field within $4\%$. The tweezer lattice constant they provided is $8.52 \mu m$. 

\begin{figure}[t]
    \centering
    \includegraphics[width=0.8\linewidth]{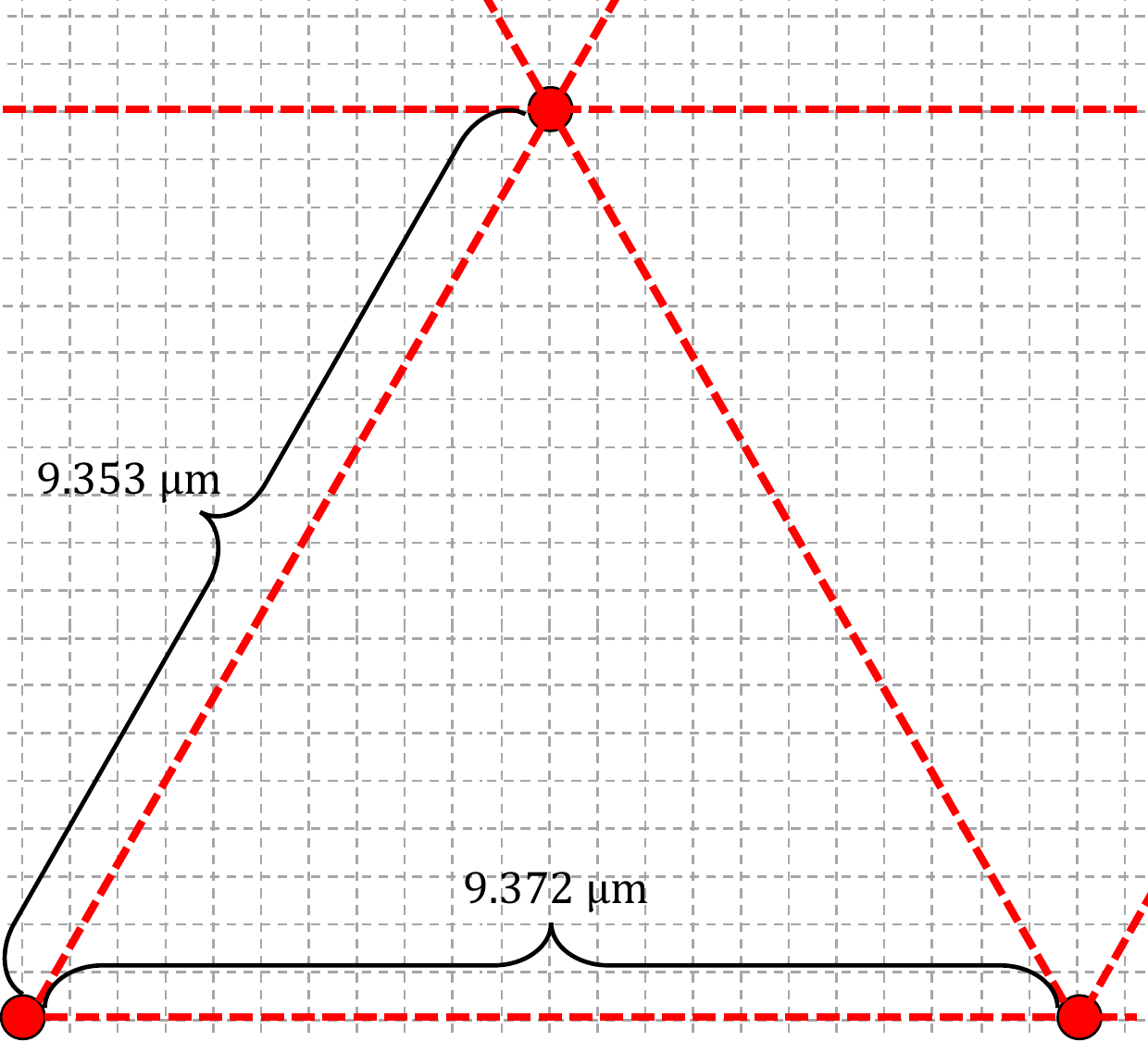}
    \caption{A schematic diagram of the unit cell of a triangular tweezer lattice arranged on a square optical lattice with a lattice constant of 
$a_\mathrm{square}=0.426\mu m$ in a cavity.}
    \label{triangular}
\end{figure}

Assuming that the coupling variance between the atoms and the cavity field does not exceed the current experimental precision when the tweezer lattice constant is above $8.52 \mu m$, we can extend this situation to a two-dimensional square optical lattice and present a scheme for experimentally realizing a tweezer triangular lattice. Here, we set the lattice constant of this square optical lattice to be the half-wavelength of the cavity beam ($a_\mathrm{square}=0.5\times852 nm=0.426 \mu m$). Thus, we can place the atoms as shown in the figure to form a tweezer triangular lattice as the Fig. \ref{triangular}. The three sides of the triangular unit have slight differences, and they are $a_x=9.372\mu m$ and $a_{\Lambda}=9.353 \mu m$, respectively. This anisotropy on the tweezer lattice will lead to the anisotropy of the Rydberg interactions $V_x/V_\Lambda=a^6_x/a^6_{\Lambda}=1.012$. However, according to the result in \cite{scipost_zhouzheng}, this slight anisotropy ($V_x/V_\Lambda<1.04$) will not disrupt the order by disorder phase.

On the other hand, we also need to consider the situation where the coupling strength $g$ is inhomogeneously distributed in space. Based on the indicators provided by the experiment, we used QMC to simulate a “disordered system” with a variance of $\pm 4 \%$ for $g$. Specifically, we generated $N_r=100$ replicas with $g$ spatially uniform-random-distributed, and performed independent QMC simulations to obtain the expected values $\langle \hat{O}\rangle_i$ of the physical quantities under the specific distribution for each replica. For a disordered system, the expectation value of the observable is obtained by implementing the replica average defined as $\langle\langle\hat{O}\rangle\rangle \triangleq \sum_i \langle \hat{O}\rangle_i/N_r$. Due to the extremely large consumption of computational resources of the disordered system, we only calculated parameters $(\tilde{\mu},g)=(0,2.00)$ and $(\tilde{\mu},g)=(0.20,2.00)$ to ensure that the main conclusions of our article still hold in the disordered system. To compare with the case without disorder $\langle \hat{O}\rangle_0$, we consider such a relative change:
\begin{equation}
    \delta_{\hat{O}}=\frac{|\langle\langle\hat{O}\rangle\rangle-\langle \hat{O}\rangle_0|}{\langle \hat{O}\rangle_0}
\end{equation}
and we find that at $(\tilde{\mu},g)=(0.20,2.00)$, $\delta_E=0.00091\% $, $\delta_{\rho_b}=0.0097\%$, $\delta_{\rho_a}=0.65\%$, $\delta_{S(\vec{Q})}=0.028 \%$, while at $(\tilde{\mu},g)=(0,2.00)$, $\delta_E=0.18\% $, $\delta_{\rho_b}=2.12\%$, $\delta_{\rho_a}=2.99\%$, $\delta_{S(\vec{Q})}=14.59 \%$. For all the physical quantities measured above, the disorder caused by the inhomogeneous $g$ is more pronounced on the first-order phase transition line than when it is not on. Following the same analysis in the effect of long-range interaction, the high degeneracy around the $Z_2$ symmetry line is strongly sensitive to a tiny energy scale, so that the disorder can bring in a large discrepancy with the pure system. However, the SRC phase is still robust against disorder.

Definitely, it does not mean the state-of-the-art experimental platform is sufficiently ready. The high-quality cavity with a large Rydberg array is still extremely challenging nowadays. However, the rapid development of the experiment always provides theoretical research with a very optimistic future. Maybe, just like the recent breakthrough on the Rydberg dressing atom in the optical lattice \cite{ebhm}, directly trapping the Rydberg dressing atom in the triangular optical lattice in the cavity will become possible in the near future. 

\subsection{superradiant solid phase}
\begin{figure}[h]
    \centering
    \includegraphics[width=0.99\linewidth]{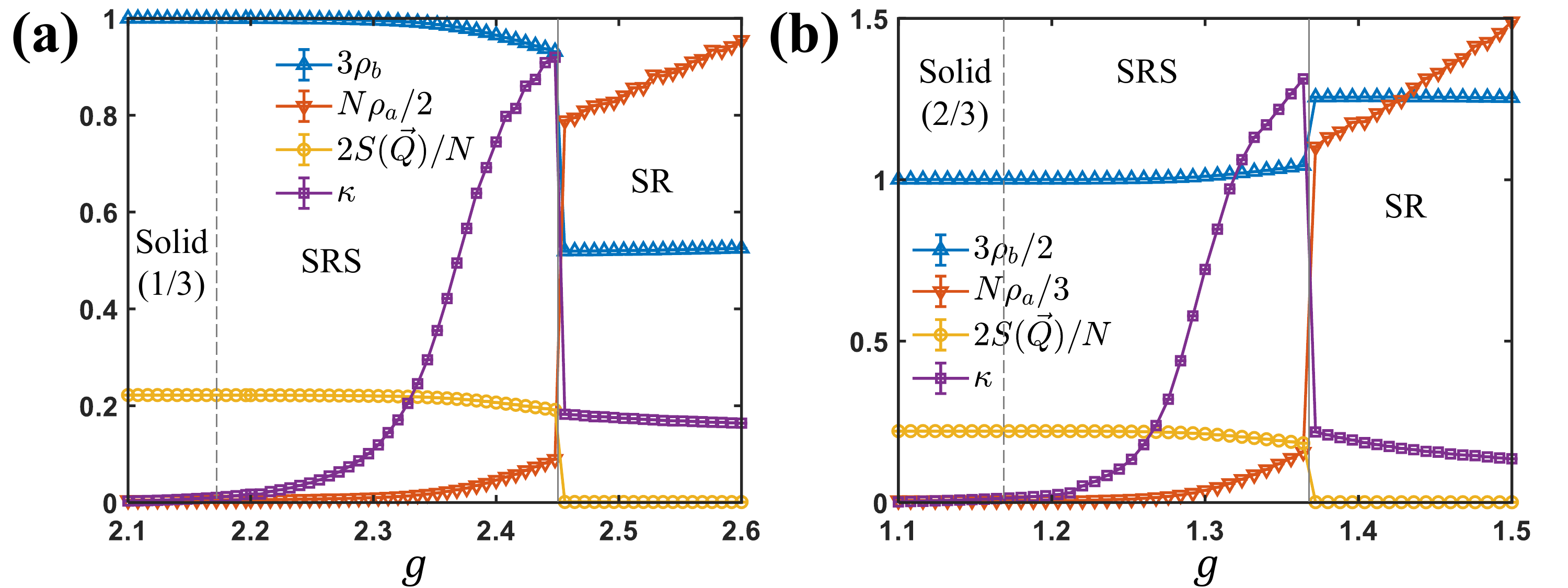}
    \caption{The photon density $\rho_a$, the Rydberg atom density $\rho_b$, the structure factor $S(\vec{Q})$ , and the compressibility $\kappa$ in SRS phases near (a) solid(1/3) phase (at $\tilde{\mu}=-2.75$) and (b) solid(2/3) phase (at $\tilde{\mu}=2.80$). The dash (solid) line marks the second (first) order QPT. The data are simulated by the QMC at $L=24$ and $\Delta=9$.}
    \label{fig:SRSs}
\end{figure}

As discussed in our previous works \cite{zhang13,ANgaoqi}, the superradiant solid (SRS) phases usually stay around the solid phase within a small region. Indeed, the previously observed SRS phase also exists in the triangular lattice, as the QMC simulation results in Fig. \ref{fig:SRSs}. The finite $S(\vec{Q})/N$ and $\rho_a$ indicate that the translational and $U(1)$ symmetries are spontaneously broken. Meanwhile, this SRS phase follows the same mechanism in the chain and square lattice systems, which is irrelevant to the OBD algorithm.

\subsection{Topological sector}
\begin{figure}[b]
    \centering
    \includegraphics[width=0.99\linewidth]{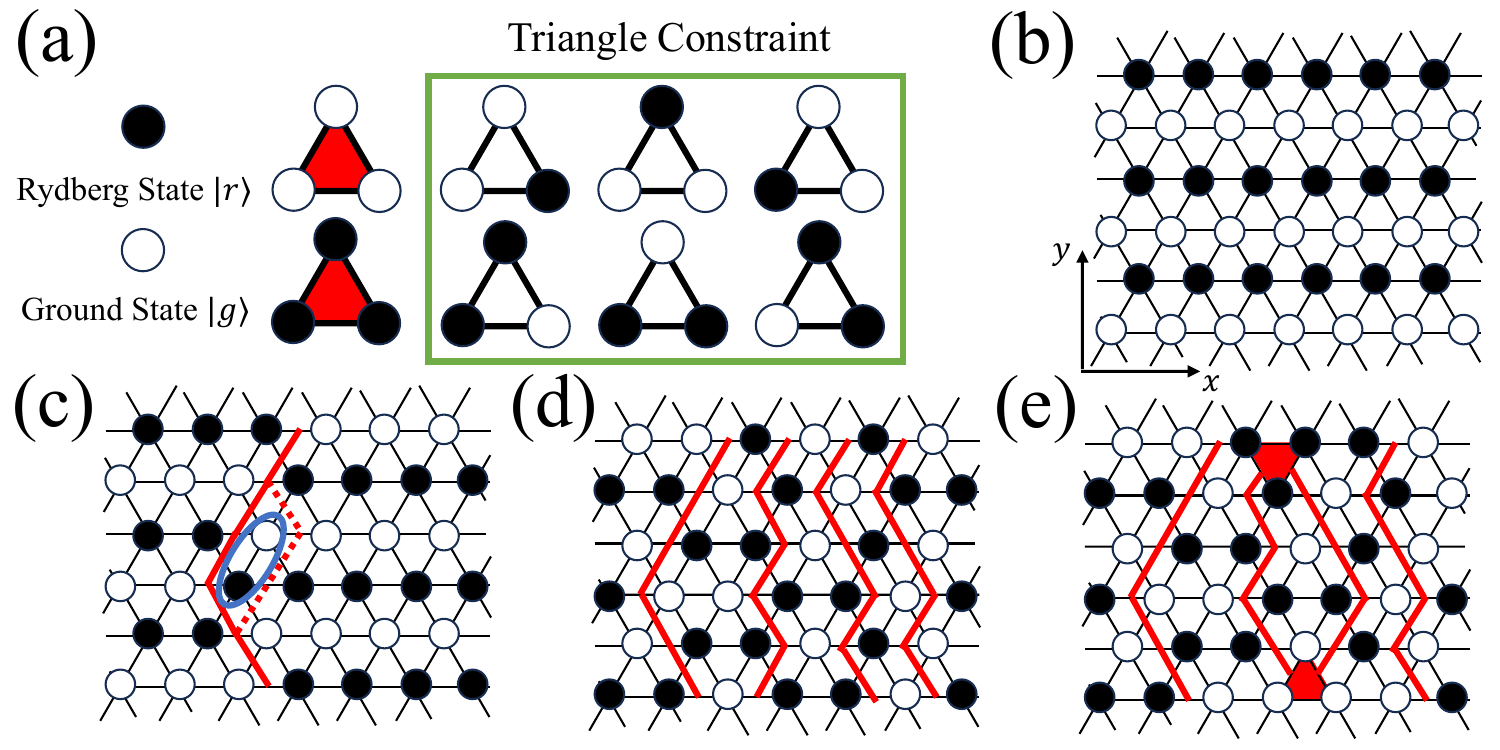}
    \caption{Schematic diagrams related to the topology. (a) Triangle constraint with fractional charge (red triangle); (b) Configuration of stripe order with winding number $N_W=0$; (c) Configuration of single domain wall (red line) which can be changed to new shape (dashed red line) without changing $N_W=1$ by local exchange interaction (blue circle); (d) Configuration of SRC phase with $N_W=2L_x/3$; (e) Possible fractional excitation at high temperature.}
    \label{fig:topo}
\end{figure}

As emphasized in the introduction, the strong Rydberg interactions and geometric frustration lead to a disordered ground state with high macroscopic degeneracy, which is related to the topological sectors. Here, we would like to make it clearer. As shown in Fig.\ref{fig:topo}(a), around the $Z_2$ symmetry line, empty and full triangles (fractional charges in lattice gauge theory) are forbidden or gaped according to the triangle constraint. Then, the restricted Hilbert space satisfying the triangle rule can be separated into different topological sectors with a different number of the degeneracy. The topological sector can be labeled by the winding numbers $N_W$ , which equals the number of antiferromagnetic bonds in each direction (here only the $x$ component is considered) \cite{Yan_annealing}. The simplest one is the stripe phase illustrated in Fig.\ref{fig:topo}(b) with $N_W=0$ and its degeneracy being one. 

Then, as demonstrated in Fig.\ref{fig:topo}(c), after altering all the configurations in the right part (by surface operator), one domain wall is introduced, and the system enters into the topological sector with $N_W=1$. We observe that exchanging the local configurations (such as spin exchanges) within the blue circle does not violate the triangular constraint. This implies that the associated topological sector remains unchanged, and consequently, the corresponding degeneracy exceeds that of the stripe phase. Because both three- and six-fold clock terms favor the same topological sector with $N_W=2L_x/3$, the winding number during the QPT between SRC and solid phase should be unchanged. Furthermore, as shown in Fig.\ref{fig:topo}(e), because the fractional charges are confined, the fractional charge pair with only a short distance can be excited at low temperature due to the quantum and thermal fluctuations.
\begin{figure*}[t]
    \centering
    \includegraphics[width=0.99\linewidth]{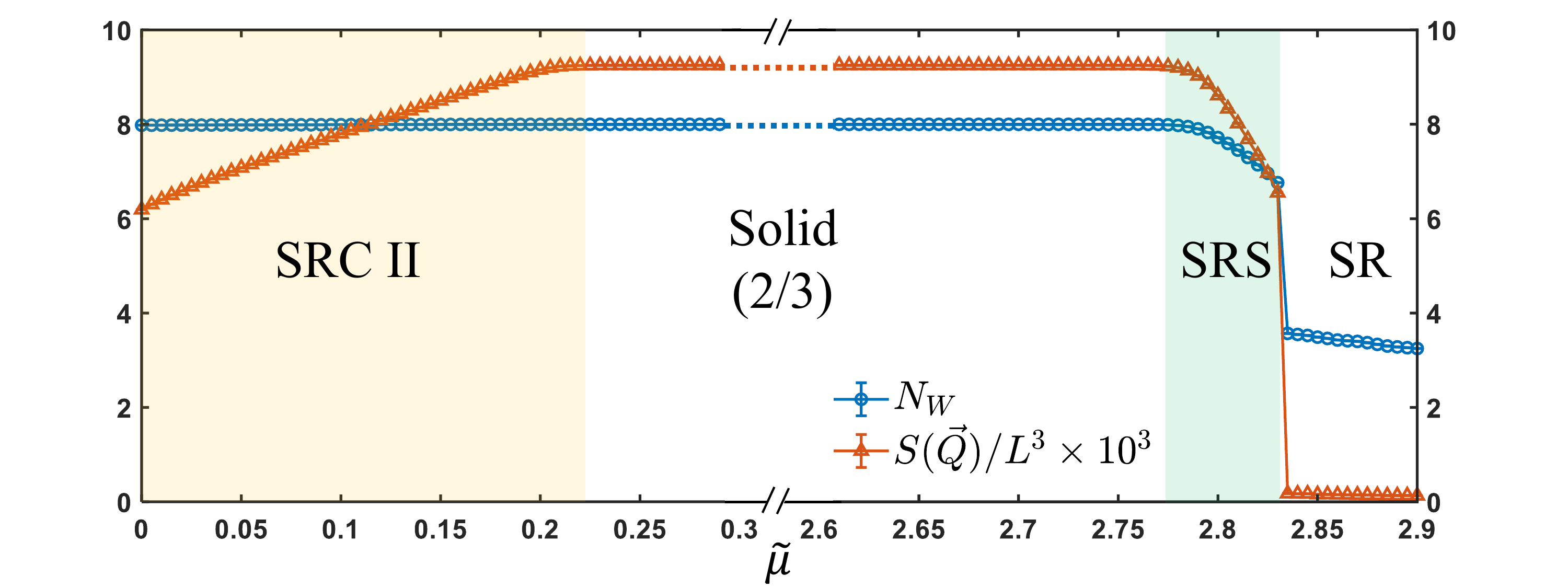}
    \caption{The winding number $N_W$ and the structure factor $S(\vec{Q})/N$ vs $\tilde{\mu}$ at $g=1.30$. The SRC phase(left) and the SRS phase(right) determined by $S(\vec{Q})/N$ are marked with yellow and green background, respectively.  The data are simulated by the QMC at $L=12$ and $\Delta=9$.}
    \label{fig:SRC-SRS}
\end{figure*}

As the comparison in Fig.\ref{fig:SRC-SRS}, the winding number during the QPT from the solid to the SRS phase is smoothly changed, which means the ground state of the SRS phase has no relation with the topological sector, otherwise a clear staircase would be expected as the TIM or XXZ model \cite{Yan_annealing,cpl_zhouzheng,scipost_zhouzheng,xxz_zhang}. Differently, as shown in the left side of Fig.\ref{fig:SRC-SRS}, the winding number during the QPT between the solid and SRC phase is almost unchanged and very close to $2L_x/3$, just the same as the prediction.

\end{document}